\documentclass[aps,amsmath,amssymb,prf]{revtex4}
\usepackage{lineno,hyperref}
\usepackage{rotating, graphicx}
\usepackage{epstopdf}
\usepackage{mathptmx}
\usepackage{amsmath}
\usepackage{caption}
\usepackage{color}

\graphicspath{{./graph/}}
\usepackage{color}

\begin{document}

\title{Dynamics of retracting surfactant-laden ligaments at intermediate $Oh$ number}

\author{Cristian R. Constante-Amores$^1$}
\author{Lyes Kahouadji$^1$}\email{l.kahouadji@imperial.ac.uk}
\author{Assen Batchvarov$^1$}
\author{Seungwon Shin$^2$}
\author{Jalel Chergui$^3$}
\author{Damir Juric$^3$}
\author{Omar K. Matar$^1$}
\affiliation{$^1$Department of Chemical Engineering, Imperial College London, South Kensington Campus, London SW7 2AZ, United Kingdom \\}
\affiliation{$^2$Department of Mechanical and System Design Engineering, Hongik University, Seoul 121-791, Republic of Korea \\}
\affiliation{$^3$Laboratoire d'Informatique pour la M\'ecanique et les Sciences de l'Ing\'enieur (LIMSI), Centre National de la Recherche Scientifique (CNRS), Universit\'e Paris Saclay, B\^at. 507, Rue du Belv\'ed\`ere, Campus Universitaire, 91405 Orsay, France}
\date{\today}

\begin{abstract}

The dynamics of ligaments retracting under the action of surface tension occurs in a multitude of natural and industrial applications; these include inkjet printing and atomisation.
We perform direct, fully three-dimensional, two-phase numerical simulations of the retracting process over a  range of system parameters that account for surfactant solubility, sorption kinetics, and Marangoni stresses. Our results indicate that the presence of surfactant inhibits the `end-pinching' mechanism and promotes neck re-opening through Marangoni-flow; this is induced by the formation of surfactant concentration gradients that drive flow-reversal towards the neck. The vortical structures associated with this flow are also analysed in detail. We also show that these Marangoni stresses lead to interfacial rigidification, observed through a reduction of the retraction velocity and ligament  kinetic energy.

\end{abstract}

\maketitle

\section{Introduction\label{intro}}
Interfacial breakup is often accompanied by the formation of satellite droplets \cite{Plateau_1873,Rayleigh_1879,Eggers_rpp_2008}. Following breakup, the interface assumes the shape of liquid threads or ligaments which undergo further capillary breakup if their lengths exceed their perimeters  \cite{Plateau_1873,Rayleigh_1879}; if not, then they retract into a single spherical drop or a multitude of droplets. Both the breakup and retraction processes are driven by surface tension in order to minimise interfacial energy. These phenomena are observed in multiple applications,  such as, atomisation or spray formation \cite{Marmottant_jfm_2004,Eggers_rpp_2008}, ink-jet printing or micro-encapsulation \cite{Eggers_rmp_1997,Basaran_aiche_2002,Hoath_pf_2013}.

In the absence of surfactant, the retracting dynamics of a Newtonian liquid thread surrounded by a passive ambient gas has been studied by \citet{Schulkes_jfm_1996}, and \citet{Notz_jfm_2004} using a Galerkin finite element approach under a two-dimensional axisymmetric assumption in the  azimuthal direction, and by also assuming another symmetry through the ligament mid-plane in the longitudinal direction;
the flow is parameterised by  the ligament aspect ratio, $L_o=L/R$, and Ohnesorge number, $Oh=\mu / \sqrt{\rho \sigma R}$, where  $L$ and $R$ are the initial half length and radius of the ligament, while $\rho$, $\mu$, and $\sigma$ are the density, dynamic viscosity, and surface tension, respectively. 

\citet{Schulkes_jfm_1996} has simulated the ligament retraction towards a minimum radius of $r \sim 0.8 R$, where $r$ is the radial coordinate, while \citet{Notz_jfm_2004}  were able to reach  $r \sim 10^{-4} \times R$. \citet{Notz_jfm_2004} have also presented the temporal evolution of a retracting ligament of aspect ratio $L_o = 15$ for three different regimes depending on the magnitude of $Oh$:  (i) low  $Oh$ values ($Oh \sim 10^{-3}$), where capillarity is more dominant than viscous forces, the outcome is the formation of two bulbous regions at both ends of the ligament that pinchoff eventually leading to the formation of a smaller, secondary ligament; (ii) intermediate $Oh$ values  ($Oh \sim 10^{-2}$), where there is a balance between viscous and capillary forces, a situation that culminates in the breakup of the ligament into three droplets; (iii) at higher $Oh$ values ($Oh \sim 10^{-1}$), which reflects the dominance of viscous forces and for which the retraction is not accompanied by breakup but by the formation of a single spherical drop.

\citet{Notz_jfm_2004} have also presented a regime map of the ligament evolution prior to its eventual breakup, varying both $L_o$ and $Oh$. \citet{Castrejon-Pita_prl_2012}  performed experiments on retracting ligaments beyond breakup, and, more recently,  \citet{Anthony_prf_2019} extended the work of \citet{Notz_jfm_2004}, considering all possible ranges of fluid properties ($Oh = 10^{-3}-10^0$) and aspect ratios ($L_o=5-10^3$). \citet{Hoepffner_jfm_2013} have shown that a cylindrical ligament undergoes capillary-driven `end-pinching' at low values of Ohnesorge number (below $Oh \sim 0.002)$. As $Oh$ increases, a viscous boundary layer is localised in the region of the neck, due to its high curvature,  which may detach leading to the formation of a jet inside the thread, reopening the neck and, subsequently, inhibiting the end-pinching mechanism. This jet is the onset of formation of a vortex ring inside the bulbous ends of the ligament.

\citet{Wang_jfm_2019} have performed both experimental and two-dimensional axisymmetric computations to determine the influence of capillary waves on the ligament stability at low viscosities (i.e., low $Oh$ numbers). They have shown that the end-pinching  mechanism always takes place for large $L_o$ (i.e., $L_o > 14$ for $Oh < 0.03$). At intermediate $L_o$ (i.e., $5 \leq L_o \leq 14$ for $Oh < 0.03$), the retraction time is too short for the neck to lead to an end-pinching mechanism, and the dynamics are induced by the interaction of the superficial capillary waves. This capillary wave interaction provides another  
mechanism to generate equal-sized droplets. Recently, \citet{Conto_sr_2019} have analysed 
the capillary retraction of Newtonian ligaments identifying three distinct regions in the system: the body of the ligament, the growing spherical bulbous end regions, and the intermediate region which connects the bulbous ends to the ligament body.

The studies summarised in the foregoing have shown that the ligament dynamics are characterised by a longitudinal retraction followed by one or possibly several breakup events \cite{Eggers_prl_1993,Eggers_rmp_1997,Day_prl_1998,Lister_pf_1998,Castrejon_pnas_2015} which lead to satellite droplet formation. The presence of surfactant, either as an additive or a contaminant, influences the interfacial dynamics via reduction of the mean surface tension, and the creation of Marangoni stresses. \citet{Craster_pof_2002} showed that insoluble surfactant retard thread pinchoff but do not alter the breakup scalings; this is because the surfactant is convected away from the thread neck by the capillary-driven flow.

\citet{Mcgough_prl_2006} observed the formation of microthreads, which connect drops during the thinning of surfactant-covered threads. This thinning towards the breakup singularity follows the scalings predicted by \citet{Eggers_prl_1993}. \citet{Kamat_prf_2018} performed both experiments and simulations  of pendant droplets covered by insoluble surfactant and showed that Marangoni stresses are responsible for microthread formation. These stresses act near the pinch point giving rise to reduced rates of thread thinning. \citet{Kamat_prf_2018} further observed that as the thread thins, a stagnation point is formed leading to surfactant accumulation at this point resisted by Marangoni-induced flow.

Surfactant solubility provides additional complexity due to its influence on Marangoni-driven transport along the interface, which influences the stability and dynamics of a number of flows driven by capillary instabilities. \citet{Liao_Lagmuir_2004} observed experimentally that the addition of a soluble surfactant at high concentrations enhances the asymmetric behaviour of a liquid bridge. \citet{Jin_pof_2006} and \citet{Jin_pof_2007} studied the detachment of a viscous drop in the presence of soluble surfactant. They have shown that solubility alters the drop dynamics in terms of neck thinning. For slow adsorption-desorption kinetics, characterised by a large, suitably-defined Biot number, $Bi$, solubility does not influence the neck dynamics since surfactant behaves as an insoluble additive leading to the formation of primary and secondary necks, and the capillary breakup of the former. For faster adsorption-desorption kinetics (i.e. increasing $Bi$), the neck dynamics can transition through different regimes such as the existence of only one neck, the thinning of the secondary neck, or the inhibition of  neck thinning.  As the Biot number increases, the mean interfacial concentration approaches its equilibrium value, as the rate of mass transfer between the bulk and the interface becomes comparable to, and eventually, exceeds the  convection rates which give rise to the formation of Marangoni stresses. \citet{Craster_jfm_2009} have studied the breakup of a viscous thread in the presence of soluble surfactant at concentrations that are potentially above 
the critical micelle concentration and have shown that Marangoni stresses cause the formation of large satellites. This prediction was observed experimentally by \citet{Kovalchuk_langmuir_2016,Kovalchuk_jcis_2018} where the satellite drop size increased by up to three times when soluble surfactant was added; the diffusion from the bulk to the interface has also been studied experimentally by Roche {\it et al}. [42]. \citet{Craster_jfm_2009} have also shown that the scalings for the minimum neck radius and axial velocity as the breakup singularity is approached are the same as the ones derived by \citet{Eggers_prl_1993} as surfactant is swept away from the thinning region.

In this paper, we perform numerical simulations of the Navier-Stokes equations coupled to transport equations for the interfacial and bulk surfactant species in order to analyse the effect of soluble surfactant on the dynamics of retracting ligaments. Our study accounts for surfactant solubility, sorption kinetics, bulk and interfacial diffusion, and Marangoni stresses. A hybrid front-tracking/level-set approach is used to resolve the interfacial dynamics \cite{Shin_jcp_2018}. We use the results of our simulations to elucidate the delicate interplay between the flow dynamics and the surfactant physico-chemical effects that underlies the mechanisms responsible for several phenomena of interest; these include ligament retraction with and without breakup, and, in the latter case, subsequent re-coalescence. In this study, we show that the surfactant concentration gradients along the interface give rise to the formation of Marangoni stresses that suppress the `end-pinching' mechanism.

The rest of this article is organized as follows. Section \ref{sec:numerical_methods} presents the governing equations for the flow and surfactant transport, the simulation configuration, and the numerical methods. In Section \ref{sec:Results}, we present a discussion of our results focusing on the effect of surfactant on the dynamics of the thread, a parametric study with respect to the governing surfactant parameters, and a detailed analysis of the vorticity. Finally, concluding remarks are provided in Section \ref{sec:conclusions}.

\section{Formulation and problem statement}\label{sec:numerical_methods}

\subsection{Governing equations}

The numerical simulations are performed by solving the two phase Navier-Stokes equations in a three-dimensional Cartesian domain $\textbf{x} = \left(x, y, z \right)$.  Surfactant transport was resolved in both the liquid bulk and on the interface by convection-diffusion equations describing the transport of surfactant species in the bulk and on the interface, with concentrations $C$ and  $\Gamma$, respectively. The source term of the momentum related to the surface tension forces is treated by using a hybrid interface-tracking/level-set method presented by \citet{Shin_jmst_2017}. The surface force is decomposed into its normal component for the normal stress jump across the interface, and its tangential component, which is associated with the surface gradient of the surface tension 
due to the presence of surfactant \cite{Shin_jcp_2018}. All variables are rendered non-dimensional by using the following scaling where the tildes designate dimensionless quantities:
\begin{equation}
\quad \tilde{r}=\frac{r}{R}, 
\quad \tilde{t}=\frac{t}{t_{_R}}, 
\quad \tilde{\textbf{u}}=\frac{\textbf{u}} {u_{_R}},
\quad \tilde{p}=\frac{p}{\rho_{_l} u_{_R}^2}, 
\quad \tilde{\rho}=\frac{\rho}{\rho_{_l}}, 
\quad \tilde{\mu}=\frac{\mu}{\mu_{_l}}, 
\quad \tilde{\sigma}=\frac{\sigma}{\sigma_s},
\quad \tilde{\Gamma}=\frac{\Gamma}{\Gamma_\infty},
\quad \tilde{C}=\frac{C}{C_{\infty}},
\quad \tilde{C}_s = \frac{C_s}{C_{\infty}},
\end{equation}
\noindent
where, $t$, $\textbf{u}$, and $p$ stand for time, velocity, and pressure, respectively. The physical parameters correspond to the liquid density $\rho_l$, and viscosity, $\mu_l$, and  the surfactant-free surface tension, $\sigma_s$. The length and time scales are normalised by the initial ligament radius $R$ and the capillary breakup time scale $t_{_R}=\sqrt{\rho_{_l} R^3/\sigma_s}$, respectively. Hence, velocities are scaled by the capillary velocity $u_{_R}=R/ t_{_R}= \sqrt{\sigma_s/(\rho_{_l} R)}$. Additionally, $\Gamma_\infty$ and $C_\infty$ are  the  interfacial concentration  at  saturation and  the  bulk  concentration, respectively. Finally, $C_s$ is the concentration of surfactant in the bulk sub-phase, immediately adjacent to the interface.  As a result of this scaling, the dimensionless equations are

\begin{equation}
 \nabla \cdot \tilde{\textbf{u}}=0,
\end{equation}
\begin{equation}\label{NS-eq}
\tilde{\rho} (\frac{\partial \tilde{\textbf{u}}}{\partial \tilde{t}}+\tilde{\textbf{u}} \cdot\nabla \tilde{\textbf{u}}) + \nabla \tilde{p}  =   Oh ~ \nabla\cdot  \left [ \tilde{\mu} (\nabla \tilde{\textbf{u}} +\nabla \tilde{\textbf{u}}^T) \right ] +
\int_{\tilde{A}\tilde{(t)}} 
(\tilde{\sigma} \tilde{\kappa} \textbf{n} +   \nabla_s  \tilde{\sigma})  \delta (\tilde{\textbf{x}} - \tilde{\textbf{x}}_{_f} )\mbox{d}\tilde{A},
\end{equation}
\begin{equation} 
\frac{\partial \tilde{C}} {\partial \tilde{t}}+\tilde{\textbf{u}}\cdot \nabla \tilde{C}= \frac{1}{Pe_b} \nabla\cdot(\nabla \tilde{C}),
\end{equation}
 \begin{equation}
 \label{interface_nd}
 \frac{\partial \tilde{\Gamma}}{\partial \tilde{t}}+\nabla_s \cdot (\tilde{\Gamma}\tilde{\textbf{u}}_t)=\frac{1}{Pe_s} \nabla^2_s \tilde{\Gamma}+ \tilde{J},
 \end{equation}
\begin{equation} 
\tilde{J}  = Bi \left ( k  \tilde{C}_s (1-\tilde{\Gamma})- \tilde{\Gamma}  \right ),
\label{eq:flux}
\end{equation}

\begin{equation} 
\tilde{\sigma}=1 + \beta_s \ln{\left(1 -\tilde{\Gamma}\right)},
\label{marangoni_eq}
\end{equation}

\noindent 
where the density and viscosity are given by $\tilde{\rho}=\rho_{_g}/\rho_{_l} + \left(1 -\rho_{_g}/\rho_{_l}\right) H\left(\tilde{\textbf{x}},\tilde{t}\right)$ and $\tilde{\mu}=\mu_{_g}/\mu_{_l}+ \left(1 -\mu_{_g}/\mu_{_l}\right) H\left( \tilde{\textbf{x}},\tilde{t}\right)$
wherein $H\left( \tilde{\textbf{x}},\tilde{t}\right)$ represents a smoothed Heaviside function, which is zero in the gas phase and unity in the liquid phase, while subscripts `$g$' and `$l$' designate the gas and liquid phase, respectively; $\kappa$ and $\bf{n}$ denote the curvature and the outward-pointing unit normal, respectively, and the surface gradient operator is given by $\nabla_s =[\bf{I}-\bf{nn}]\cdot \nabla$ wherein $\bf{I}$ is the identity tensor; $\tilde{\textbf{x}}_{_f}$ is a parameterization of the interface $\tilde{A}(\tilde{t})$, 
and $\delta(\tilde{\textbf{x}}-\tilde{\textbf{x}}_{_f})$ is a Dirac delta function that is non-zero only when $\tilde{\textbf{x}}=\tilde{\textbf{x}}_{_f}$.

Additionally, $\tilde{\mathbf{u}}_t= \left ( \tilde{\mathbf{u}}_s \cdot \mathbf{t} \right ) \mathbf{t}$ is the tangential velocity on the interface in which $\tilde{\mathbf{u}}_s$ represents the surface velocity; $\tilde{J}$ is the sorptive flux, which provides a relationship between $\tilde{C}$ and $\tilde{\Gamma}$ that connects the bulk and interfacial concentrations. The left-hand-side of Eq. (\ref{interface_nd}) represents the transient and convective transport of surfactant at the interface, and its right-hand side models interfacial diffusion and bulk-interface mass exchange.

The dimensionless parameters appearing in equations (\ref{NS-eq}-\ref{marangoni_eq}) are given by
\begin{equation}\label{dimless}
\quad Oh=\frac{\mu_{_l}}{\sqrt{\rho_{_{l}} \sigma_s R}}, ~~ 
\quad Bi=\frac{k_d R}{u_{_R}}, ~~ 
\quad \beta_s= \frac{\Re T \Gamma_\infty }{\sigma_s },~~ 
\quad Pe_s=\frac{ u_{_R} R}{D_s},~~
\quad Pe_b=\frac{ u_{_R} R}{D_b},
\end{equation}
where $Oh$ denotes the Ohnesorge number and measures the relative importance of viscous to surface tension forces, $Bi$ is the Biot number representing the ratio of characteristic desorptive to convective time-scales, $\beta_s$ is the elasticity number, which measures the sensitivity of the surface tension to changes in surfactant interfacial concentration; $Pe_s$ and $Pe_b$ are the interfacial and bulk Peclet numbers and compare the ratio of convective to diffusive time-scales in the plane of the interface and the bulk, respectively. Finally, $k_d$ refers to the surfactant desorption coefficient, $\Re$ the ideal gas constant, $T$ the temperature, $D_s$ and $D_b$ the surfactant interfacial and bulk diffusivities, respectively.

At equilibrium, Eq. (\ref{eq:flux}) reduces to the Langmuir adsorption isotherm
\begin{equation} 
\chi=\frac{\Gamma_{eq}}{\Gamma_\infty}=\frac{k}{1+k} \mbox{,  }  \thinspace    \thinspace  k=\frac{k_a C_\infty }{k_d},
\end{equation}
where $\chi$ stands for the fraction of surface covered by adsorbed surfactant, $k$ is the adsorption parameter, which represents the ratio of adsorption to desorption time scales. Here, $k_a$ refers to the adsorption coefficient. The equation of state describing the variation of the surface tension as a function of the local interfacial surfactant concentration is given by the Langmuir relation shown in Eq. (\ref{marangoni_eq}). Marangoni stress, $\tau$, arises due to the surface tension gradients and they can be expressed in terms of gradients in $\Gamma$ as follows:
 \begin{equation}
 \label{marangoni}
 \tilde{\tau} \equiv  \nabla_s \tilde{\sigma} \cdot  \textbf{t} =-\frac{\beta_s}{1 -\tilde{\Gamma}} \left ( \textbf{t} \cdot \nabla_s\tilde{\Gamma} \right ).
\end{equation}
\begin{figure}
\includegraphics[width=0.8\linewidth]{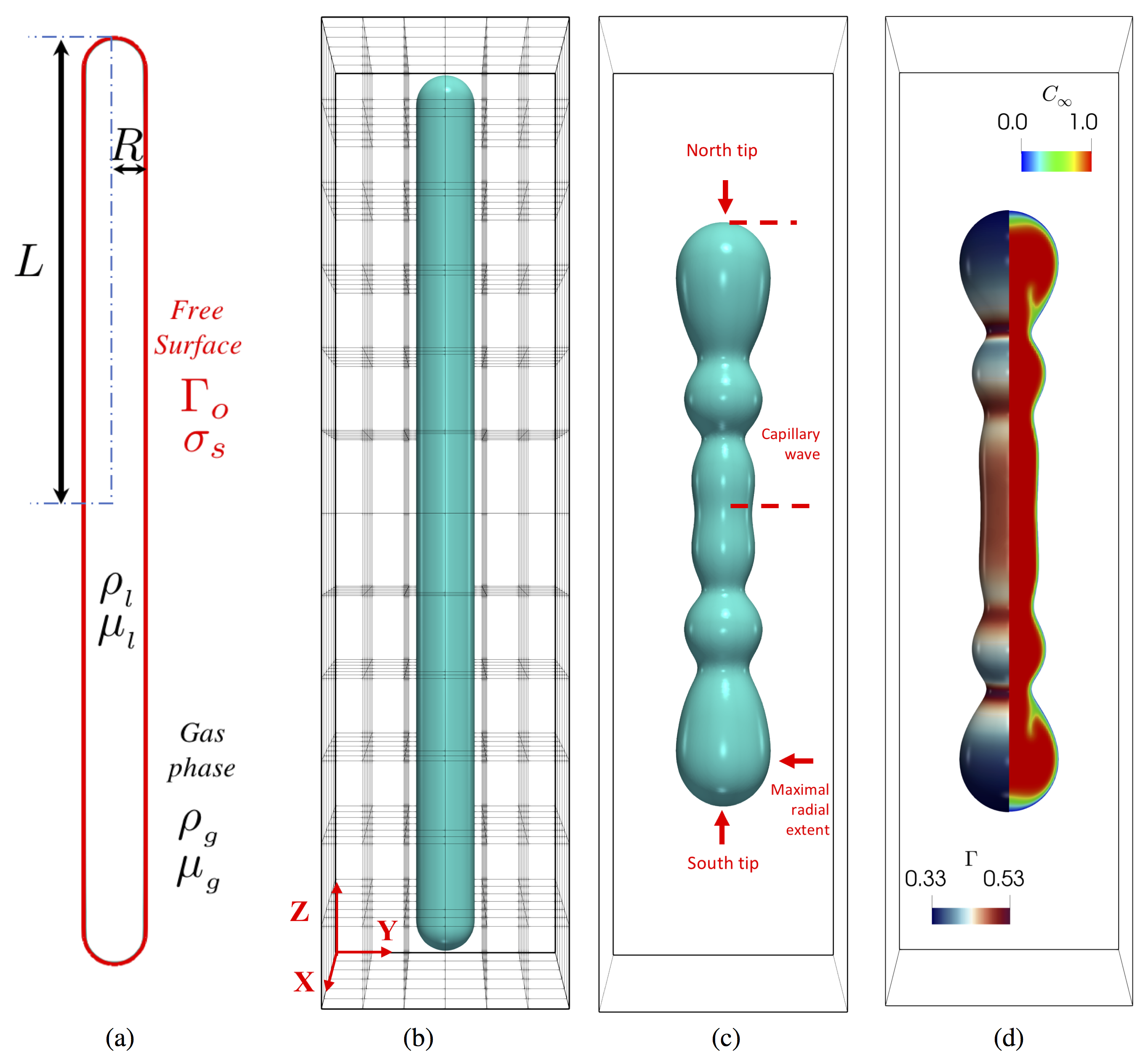}  
\caption{Schematic representation of the initial shape of the ligament, (a), highlighting  the fluid quantities and non-dimensional parameters of the system, (b), initial shape of the ligament with the computational domain of size $8R\times 8R \times 32R$ in a three-dimensional Cartesian domain $\textbf{x} = (x, y, z)$ and divided into $6 \times 6 \times12$ sub-domains; the Cartesian resolution is set to $32\times 32 \times 64$ per sub-domain, and the global resolution is $192 \times 192 \times 768$. Typical interfacial shape for a surfactant-free case at $t=7$ for $L_o=15$ and $Oh=10^{-2}$, (c), with definitions of particular locations and features whose dynamics will be discussed in the present work. (d) Typical interfacial shape for a surfactant-laden case at $t=7$ with $L_o=15$, $Oh=10^{-2}$, $Bi=1$, $Pe_s=10$, and $\chi=0.9$ with colour bars indicating the magnitude of the surfactant interfacial and bulk concentrations in the left and right halves of the ligament, respectively.}
\label{fig:intro} 
\end{figure}
\noindent For simplicity, the tildes are dropped henceforth. 

The viscosity and density ratios, $\rho_{_g}/\rho_{_l}$ and  $\mu_{_g}/\mu_{_l}$, are set to $1.2 \times 10^{-3}$ and $1.8 \times 10^{-2}$, respectively, corresponding to values for a water ligament in air. The time scale associated with the  Marangoni flow is determined from a balance between Marangoni stresses and viscous retardation, $\tau \sim \Delta\sigma/R \sim \mu_{_l} /t_{_M}$, hence $t_{_M} = \mu_{_l} R/ \Delta \sigma$, and is of order $10^{-3}$ s. However, the capillary breakup time is of order $ 10^{-2}$ s, and the time scale associated with the retraction of the ligament is also of order $ 10^{-2}$ s. For the soluble cases, we consider the properties of n-alcohols (such as n-propanol, n-butanol and n-pentanol) or dicarboxylic acid type (such as adipic and pimelic acid) as surfactants, which are characterised by desorptive time scales of $ 10^{-2}$ s \cite{Bleys_jpc_1985,Joos_jcp_1982,Joos_jcis_1989}. Therefore, for both soluble and insoluble surfactant configurations, Marangoni stress is expected to play a major role in the ligament retraction dynamics. 

\begin{figure}
\includegraphics[width=1\linewidth]{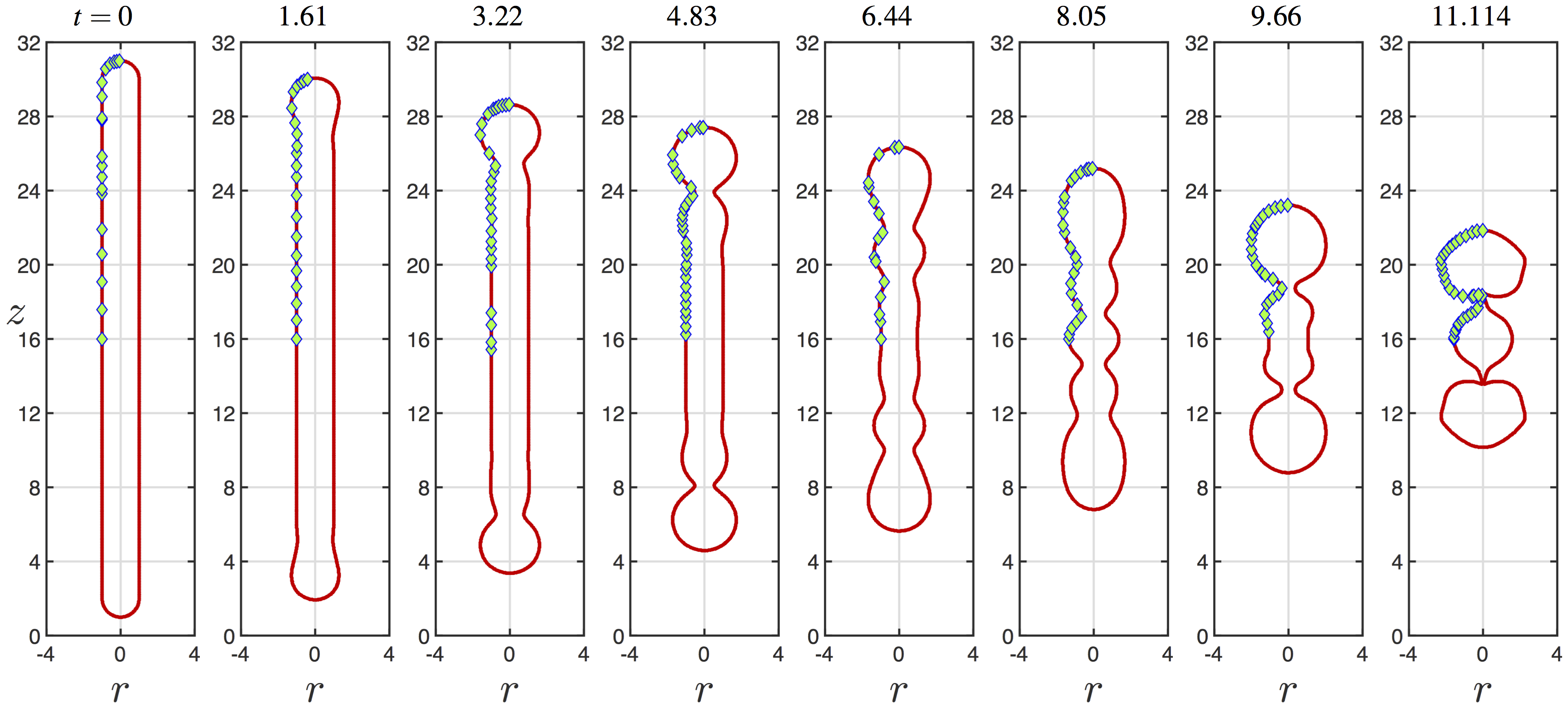}  
\caption{Spatio-temporal evolution of a retracting ligament for $L_o=15$ and $Oh=10^{-2}$. The solid lines correspond to the results of the present study and the diamonds are collected from \citet{Notz_jfm_2004}.}
\label{fig:validation} 
\end{figure}

\begin{figure}
\includegraphics[width=1\linewidth]{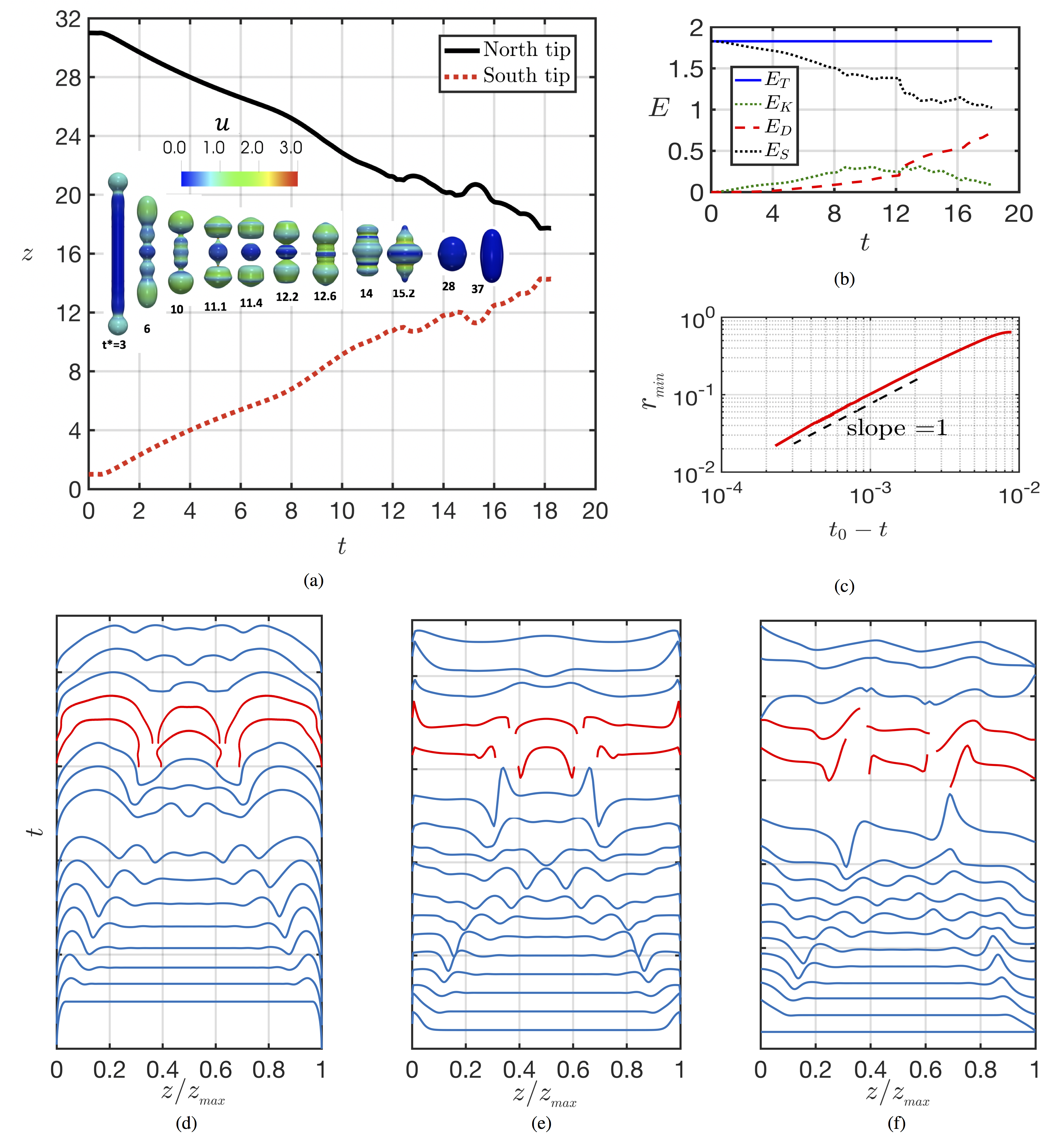}  
\caption{Surfactant-free ligament retraction for  $L_o=15$ and $Oh=10^{-2}$: (a) Temporal evolution of the location of both ligament tips, and a three-dimensional representation of the interface for the dimensionless times shown in the panel in which the colour bar depicts the velocity magnitude; (b) temporal evolution of the total, kinetic and surface energies, $E_T$, $E_K$ and $E_S$, respectively, and the energy dissipated, $E_D$; (c) scaling of the minimum radius with respect to the pinchoff time $t_o=11.115$, which agrees  with that predicted by inertio-viscous scaling theory of \citet{Eggers_prl_1993};
 (d)-(f) time-space plots of the interface, $p$, and $u_z$, respectively, with snapshots shown between $t=0-13.5$, with time intervals of $1$ between $t=0-11$ and $0.5$ between $t=11-13.5$; here, the red profiles are associated with $t=11.5$ and $t=12$, the instances at which pinchoff occurs.} 
 \label{fig:clean_case}
\end{figure}

\subsection{Problem statement and validation}

As introduced in the beginning of this section, the liquid ligament is initialised as a cylindrical thread of length $2L$ with an aspect ratio $L_o=L/R=15$ and hemispherical caps at its two ends, as shown in Fig. \ref{fig:intro}-(a). The size of the three-dimensional computation domain is  $8R \times 8R \times 32R$ where the $z$-coordinate is aligned with the height of the ligament while $x$ and $y$ are along its width. Hence a radial component is defined as $r=\sqrt{\left(x-x_o\right)^2 + \left(y-y_o\right)^2}$, where $x_o$ and $y_o$ are the abscissa and ordinate ligament position, respectively. The simulation is initialised with fluids at rest in the absence of gravity. A no-slip boundary condition is imposed on the fluid velocity at the walls of the computational domain  and a Neumann condition is imposed for the pressure $\left(\partial {p}/\partial \textbf{n}=0\right)$, where $\textbf{n}$ here refers to the normal vector at the boundaries of the computational domain. At the free surface, we set 
\begin{equation}
\textbf{n}\cdot\nabla \tilde{C} =-Pe_b Bi ( k  \tilde{C_s} (1-\tilde{\Gamma})- \tilde{\Gamma} ), 
\end{equation}
as a condition on  $\tilde{C}$. The entire domain is discretised into an Eulerian fixed regular mesh. The surfactant-free simulation was carried out on three different meshes, referred to here as M1, M2, and M3, to ensure mesh-independent results. The results presented in the study correspond to the  M2 mesh, unless stated otherwise. The code was benchmarked against the ligament profiles of \citet{Notz_jfm_2004} (see Fig. \ref{fig:validation}) on the dynamics of the surfactant-free simulation and the pichoff time. The validation of the surfactant solver has been presented previously by \citet{Shin_jcp_2018}, where the authors ensured mass conservation of the surfactant. More information of the validation, mesh refinement studies, and numerical method can be found in the Appendix.

\section{Results\label{sec:Results}}

In this section, we present a discussion of our  results, beginning by comparing our predictions with previous work on retracting surfactant-free ligaments. This serves the purpose of validating our numerical method and provides a benchmark against which to highlight the effects associated with the presence of surfactant.

\subsection{Surfactant-free ligament retraction and capillary-breakup}
We study the retraction of a surfactant-free ligament with $L_o=15$ and $Oh=10^{-2}$ previously examined by \citet{Schulkes_jfm_1996} and \citet{Notz_jfm_2004} paying particular attention to interfacial breakup and post-pinchoff dynamics. In order to inspire confidence in the reliability of the numerical method used to carry out the computations, we show in Fig. \ref{fig:validation} a comparison between our numerical predictions and those from \citet{Notz_jfm_2004} which reveals a good agreement. Figure \ref{fig:clean_case} depicts time-space plots of the interface, pressure, and the axial velocity; for the latter two, spatial variations are shown with respect to the ligament centreline. The initially motionless, cylindrical ligament undergoes retraction due to the pressure gradient between the two bulbous ends and the rest of its body, which drives flow from  these regions towards its centre in the form of a capillary wave; this dominates the early stages of the dynamics.

The retraction velocity results from a force balance between capillary and inertial forces, the latter being proportional to the rate of change of momentum of the bulbous ends. Extending the Taylor-Culick expression for the retraction of a two-dimensional  axisymmetric planar liquid sheet to a cylindrical thread, we arrive at $V=\left(2 \sigma_s/ \rho_{_l} \pi R\right)^{1/2}$, which gives a value for the vertical velocity of retraction $0.079$ m/s. This is in good agreement with the retraction velocity predicted by our simulations, $\sim 0.0789$  m/s and provides further validation of the accuracy and reliability of our numerical method.

The retraction motion and associated capillary waves form neck regions near the two ends of the ligament connecting the bulbous regions with the rest of the ligament. The pressure under these neck regions is large and drives flow away from them on time scales shorter than those related to retraction, dominating the intermediate stage of the dynamics, and promoting further necking and an even larger pressure gradient that eventually leads to a double pinchoff event for the set of parameters used to generate these results; the profiles for the interface, pressure, and axial velocity associated with this event are highlighted in red in Fig. \ref{fig:clean_case}-(d) and (f). 

Figure \ref{fig:clean_case}-(a), which highlights the temporal evolution of the north and south tips of the bulbous ligament ends during retraction, also shows that the pinchoff, which takes places at $t\sim11.11$, is followed by
the formation of three droplets (see Fig. \ref{fig:clean_case}-(a) for $t=11.4$). 
These droplets are sufficiently close that a double coalescence takes place simultaneously at  $t=12.1$ generating capillary waves that travel up and down the ligament (see Fig. \ref{fig:clean_case}-(a) for $t=12.6$, $14$, and $15.2$). These waves decelerate giving way to decaying oscillations between a spherical and an ellipsoidal ligament shape that are the main features of the late-time dynamics (see Fig. \ref{fig:clean_case}-(a) for $t=28$ and $37$).

It is also instructive to perform an analysis of the temporal variation of the system energy. The total energy $E_T$ must be constant over time and its constituents are the surface energy, $E_s= S \sigma_s$, where $S$ is the superficial area of the ligament, the kinetic energy, $E_k=\int_V (\rho \textbf{u}^2)/2 dv$, and the energy dissipated, $E_{D}=-\int_V (\tilde{\tau} : \nabla \mathbf{u}) dv$, where $\tilde{\tau}$ refers to a viscous stress tensor. As highlighted in Fig. \ref{fig:clean_case}-(b), all energies are normalised by the surface energy of a motionless spherical droplet with a volume similar to that of the ligament of aspect ratio $L_o=15$. Initially, the total energy is solely represented by the surface energy $E_s$. When the ligament retracts, part of the surface energy is transferred into kinetic energy. During the coalescence of the three droplets (at $t \sim 12.1$) the total area of the system changes significantly and a fraction of the surface energy is
dissipated (see Fig. \ref{fig:clean_case}-(b)). At longer times, $E_s \rightarrow 1$ and $E_k \rightarrow 0$, as the ligament tends towards a steady, spherical shape.  

\begin{figure}
\includegraphics[width=1\linewidth]{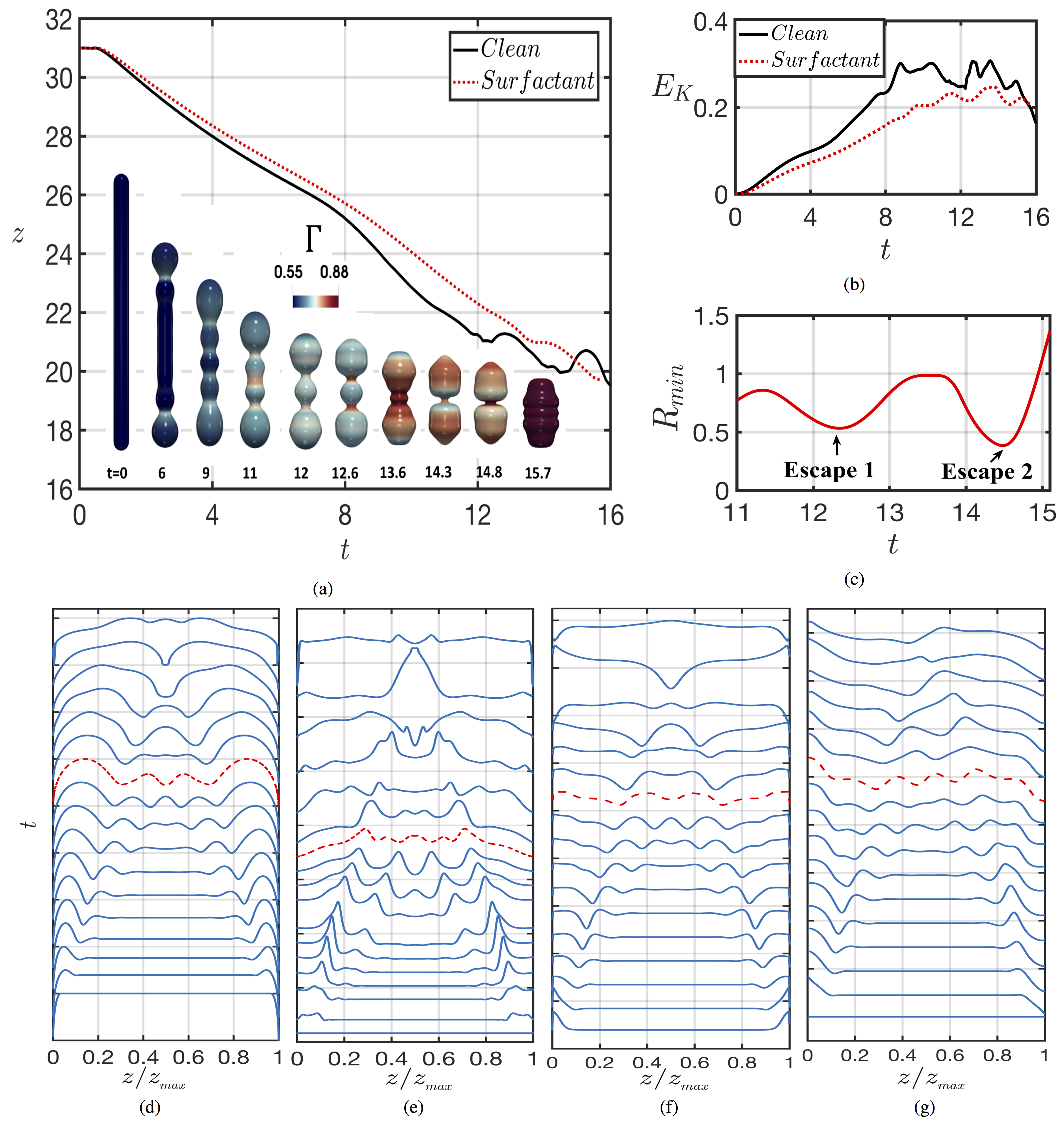}  
\caption{Ligament retraction with an insoluble surfactant for $L_o=15$, $Oh=10^{-2}$, $Pe_s=10$, $\beta_s =0.3$ and $\Gamma_o =\Gamma_\infty/2$: (a) temporal evolution of the north tip location for the surfactant-free and surfactant-laden cases, and a three-dimensional representation of the interface for the dimensionless times shown in the panel in which the colour bar depicts the magnitude of the surfactant interfacial concentration, $\Gamma$; (b) temporal evolution of the Kinetic energy $E_K$ for the surfactant-free and surfactant-laden cases; (c) temporal evolution of the neck radius highlighting the two escapes of pinchoff; (d)-(g) time-space plots of the interface, $\Gamma$, $p$, $u_z$ with snapshots shown between 
$t=0-16$ at equal time intervals; here, the red profiles are associated with $t=10$.\label{fig:surfactant_base_case}}
\end{figure}
%
%
%
\subsection{Surfactant-laden ligament retraction: escape from capillary-breakup}
In this section, we present the effect of insoluble surfactant on the dynamics of a retracting ligament with $L_o=15$, $Oh=10^{-2}$, $Pe_s=10$, $\beta_s =0.3$, and $\Gamma_o =\Gamma_\infty/2$. Figure \ref{fig:surfactant_base_case} depicts the spatio-temporal evolution of the interface and $\Gamma$ 
together with the pressure and axial velocity along the ligament centreline. 
Similar to the surfactant-free case, retraction is accompanied by the formation of capillary waves 
that dominate the dynamics leading to the collapse of the initially-cylindrical ligament towards a spherical one. The surfactant concentration $\Gamma$, which is coupled to the interfacial dynamics through the dependence of $\sigma$ on $\Gamma$, is redistributed along the interface, and achieves a maximal value around $t \sim 15$ since the ligament area decreases as it approaches a spherical shape. As shown clearly in Fig. \ref{fig:surfactant_base_case}-(a) and (b), the presence of surfactant retards ligament retraction as evidenced by the slower temporal evolution of the ligament tips and lower kinetic energy in comparison to the surfactant-free case; the retraction speed is $\sim 0.070$ m/s as compared to $\sim 0.0789$ m/s  in the `clean' case.
This is due to the surfactant-induced interfacial rigidification brought about by the Marangoni stresses, which, in turn, are caused by gradients in $\Gamma$.

\begin{figure}
\includegraphics[width=1\linewidth]{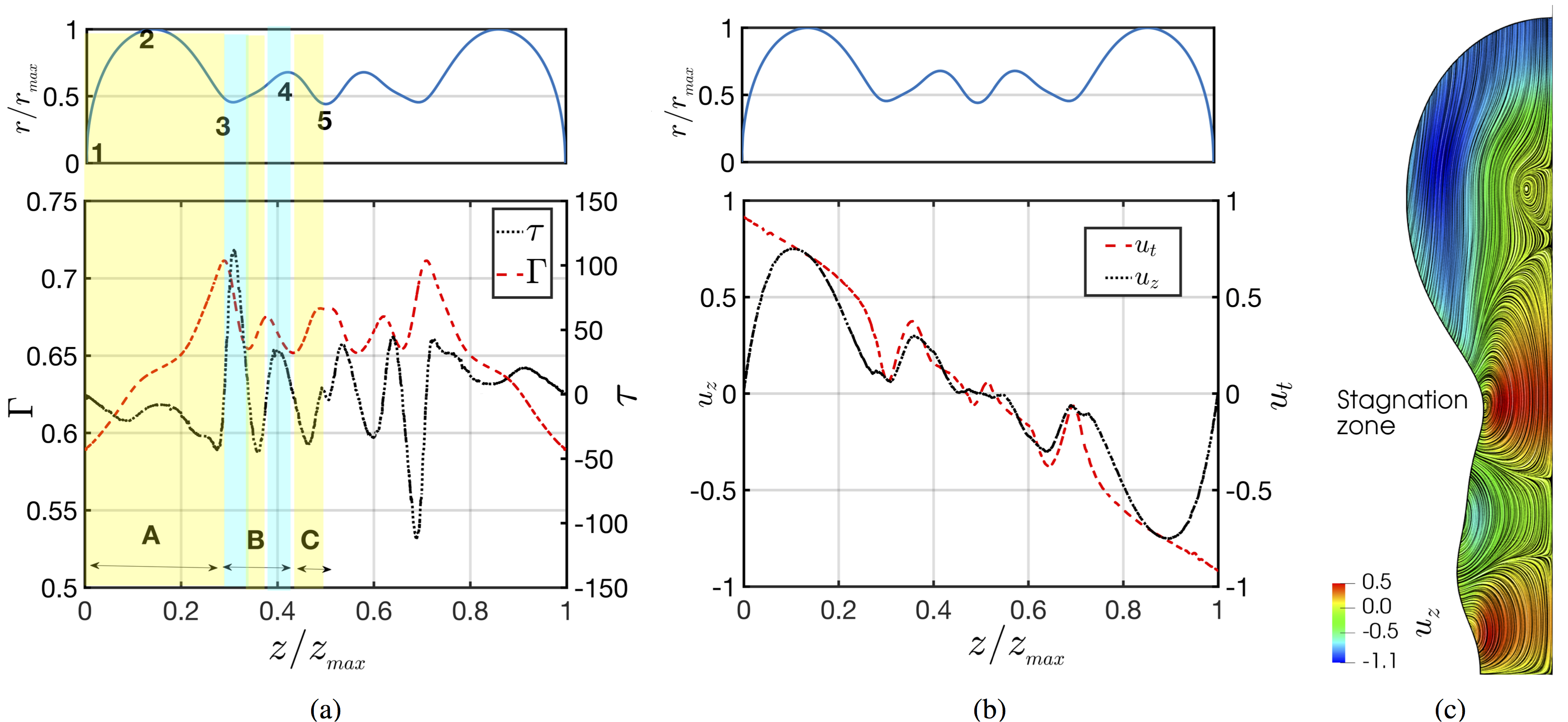}  
\caption{Spatial variation of the interfacial shape, $\Gamma$, and $\tau$, (a), the tangential interfacial velocity, $u_t$, for the surfactant-free and surfactant-laden cases, (b), and the streamline structure within the retracting ligament, (c); the parameter values are the same as in Fig. \ref{fig:surfactant_base_case} with $t=10$.}
\label{fig:surfactant_base_case_fields}
\end{figure}
In order to elucidate the coupling between interface and surfactant concentration,  we consider the interface, $\Gamma$, $p$, and $u_z$, at  $t=10$ shown in red dashed lines in panels (c)-(f) of Fig. \ref{fig:surfactant_base_case}.  
The retraction capillary waves are characterised by regions of radially-diverging and converging motion and associated higher and lower interfacial areas and therefore reduced (i.e. increased) and increased (i.e. reduced) $\Gamma$ ($\sigma$) locally, respectively.  These concentration gradients lead to Marangoni stresses that drive flow from the higher-tension radially-diverging to the lower-tension converging regions, which act to retard the interfacial motion. To further illustrate the retarding effect of the Marangoni stresses,  three distinct regions are also highlighted, as shown in Fig. \ref{fig:surfactant_base_case_fields}-(a). In Region `A', the interfacial flow diverges away from point `1', at the ligament tip, driving surfactant away from this location leading to the lowest $\Gamma$ value along the interface. There is an overall increase in $\Gamma$ from the tip towards the centre reaching a maximal value at location `3' where the interface exhibits a local minimum. The $\Gamma$ profile then undergoes oscillations in response to the wavy shape of the interface, with a local minimum and maximum in $\Gamma$ at locations `4' and `5' that coincide with a local interfacial maximum and minimum, respectively. It is clear from   Fig. \ref{fig:surfactant_base_case_fields}-(a) that $\tau < 0$ in Region `A' suggesting that the direction of the Marangoni flow is towards the ligament tip, which acts to retard the capillary-driven flow from the tip towards the centre; this retarding effect in Region `A' manifests itself through a decrease in the tangential velocity along the interface, $u_t$, as shown in Fig. \ref{fig:surfactant_base_case_fields}-(b). In Region `B', $\tau > 0$, thus Marangoni-driven flow is towards the ligament centre, which is counter to the capillary flow away from this necking region. As also indicated in Fig. \ref{fig:surfactant_base_case_fields}-(b), $u_t$, which was negative in Region `B' in the surfactant-free, becomes positive in the surfactant-laden case. 

It is also evident that using similar mechanisms, the Marangoni-driven flow reduces substantially the magnitude of $u_t$ in Region `C'. Figure \ref{fig:surfactant_base_case_fields}-(c) shows the structure of the streamlines, which characterise the flow within the ligament. It is clearly seen that the formation of several stagnation points occurs along the interface reflecting the competition between the capillary- and Marangoni-driven flows an example of which is provided by the stagnation point close to the neck region. The formation of the stagnation zone is similar to what \citet{Kamat_prf_2018} have described, reporting the deceleration of the fluid by the action of the Marangoni stress during the thinning of the fluid thread. This force competition reverses the flow direction, leading to the genesis of the stagnation zone.

\begin{figure}
\includegraphics[width=1\linewidth]{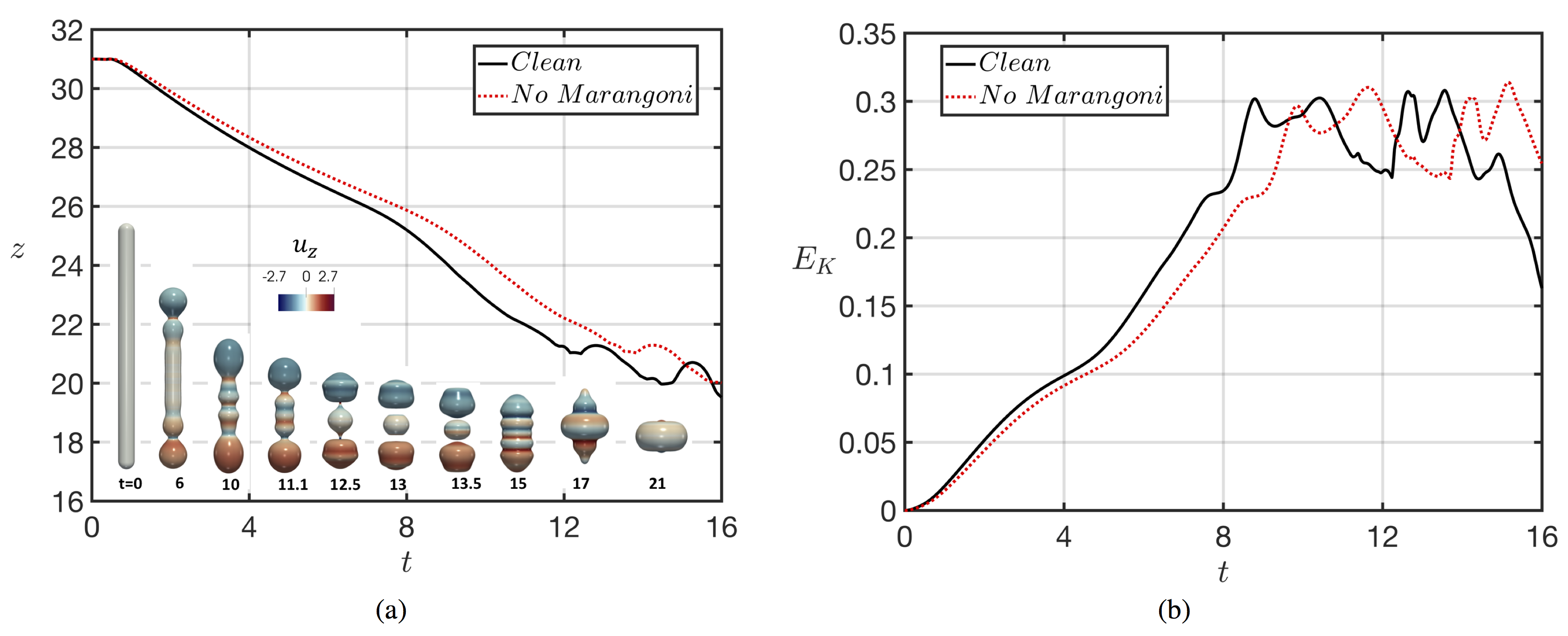}  
\caption{Dynamics of a retracting, surfactant-covered ligament with isolating the Marangoni effect: (a) temporal evolution of the north tip location for the surfactant-free and surfactant-laden cases, and a three-dimensional representation of the interface for the dimensionless times shown in the panel in which the colour bar depicts the magnitude of the surfactant interfacial concentration, $\Gamma$; (b) temporal evolution of the kinetic energy $E_K$ for the surfactant-free and surfactant-laden cases.}  
\label{fig:non_Marangoni}
\end{figure}

In order to separate the effects of mean surface tension and Marangoni stresses induced by surface tension gradients, we consider a case in which the surface tension value is reduced and given by Eq. (\ref{marangoni_eq}) using the initial interfacial concentration,  but where  no Marangoni stresses are supported. Figure \ref{fig:non_Marangoni} shows that the reduction in surface tension leads to a delay in the ligament retraction but does not prevent breakup; very similar behaviour to the surfactant-free case is observed in terms of the formation of three droplets, which eventually coalesce, and temporal evolution of the kinetic energy which undergoes a slightly delayed rise due to the slower capillary-driven flow, as expected. These results demonstrate that the prevention of the breakup is due to the  formation of Marangoni stresses rather than the reduction in surface tension. 

In Fig. \ref{fig:surfactant_base_case}-(c), we observe two escapes from breakup at $t\sim 12.3$ and $t \sim 14.3$. The interfacial shape prior to the first escape resembles that of the surfactant-free case before its breakup, shown in Fig. \ref{fig:clean_case}-(a). Hence, we will compare the flow fields of theses two cases to determine the effect of surfactant on the onset of suppression 
of the capillary breakup for the surfactant-laden case.

\begin{figure}
\includegraphics[width=1\linewidth]{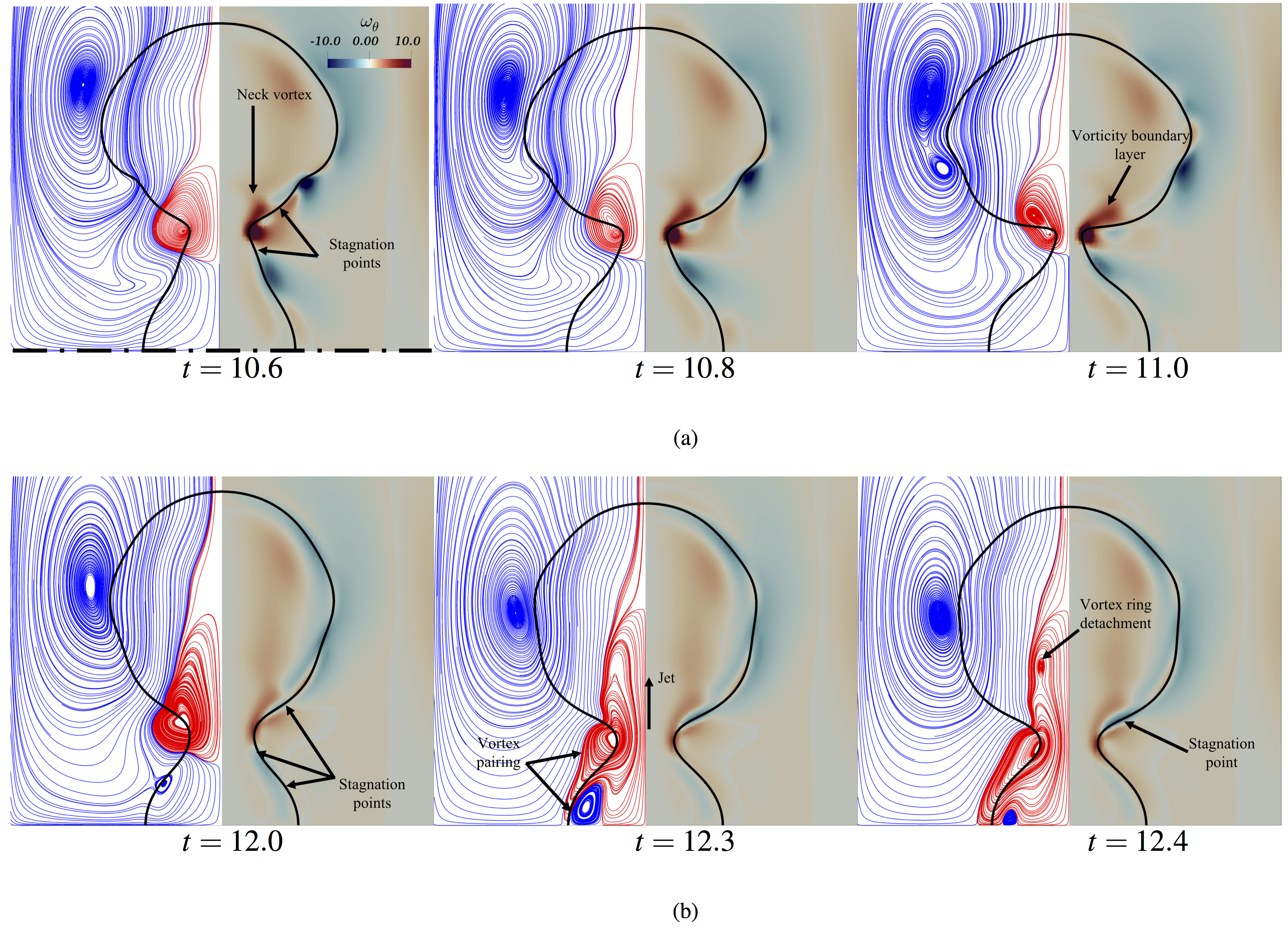}  
\caption{ \label{fig:escape_vorticity} Illustration of the surfactant-mediated mechanism underlying the first escape from pinchoff in Fig. \ref{fig:surfactant_base_case}-(c) depicting a three-dimensional representation of the velocity vector field, together with $\omega_{\theta}$ and streamlines for the surfactant-free case at $t=10.6$, 10.8, and 11, (a), and the surfactant laden base case, (b) at $t=12.0$, 12.3, and 12.4, with the same parameters as in Fig. \ref{fig:surfactant_base_case}. The framed regions of the ligament show a magnified view of the flow direction at every snapshot.}
\end{figure}

The first row of Fig. \ref{fig:escape_vorticity}-(a) shows the flow behaviour through the azimuthal vorticity, $\omega_\theta$, the instantaneous streamlines, and a three-dimensional representation of the velocity vector field for the surfactant-free case before pinchoff. Two stagnation points shown in Fig. \ref{fig:escape_vorticity}-(a) at $t=10.6$ are also observed on either side of the neck, for which there is a change in direction of the flow rotation. By inspecting $\omega_\theta$, the highest vorticity production is located
on the neck of the ligament promoted by the large interfacial curvature in that region.
A thin vorticity boundary layer is detached from the neck of the ligament and diffuses into the bulk of the bulbous region (see Fig. \ref{fig:escape_vorticity}-(a) at $t=11.0$). As the interfacial singularity is approached, the axial velocity component through the neck increases due to the associated rise in the capillary pressure, which pumps liquid rapidly away from the neck towards the bulbous region. The velocity achieves its maximum value at the moment of singularity as can be seen in Fig. \ref{fig:escape_vorticity}-(a).

The second row of Fig. \ref{fig:escape_vorticity}-(b) shows the flow behaviour for the surfactant-laden case prior to the first escape from pinchoff. Inspection of $\omega_\theta$ reveals the formation of a vortex ring also located at the neck, as shown at $t=12.0$. However, the vorticity production is not as strong as in the surfactant-free case due to the rigidification of the interface brought about by  the presence of Marangoni stresses. In comparison to the surfactant-free case, an additional stagnation point between the neck and the center of the ligament is observed as shown in Fig. \ref{fig:escape_vorticity}-(b) at $t=12.0$. The presence of this
stagnation point has an associated vortex ring which interacts with the interface. As time increases, the mutual interaction of the two vortices leads to a vortex-pairing process, as shown at $t=12.3$. This pairing up reverses the flow direction towards the neck ultimately leading to its reopening. The framed regions of the ligament show a magnified view of the flow direction and the flow reversal towards the neck. As the flow re-enters through the neck, the formation of a
jet towards the bulbous region is observed, which gives rise to a vortex ring 
that eventually detaches towards the centre of the bulbous region, as shown at $t=12.4$. This behaviour is similar to the phenomenon explained by \citet{Hoepffner_jfm_2013} (see their Fig. 6) where they showed that the formation of a vortex ring plays the primary role in the escape from breakup of viscous ligaments in the absence of surfactant.

\begin{figure}
\includegraphics[width=1\linewidth]{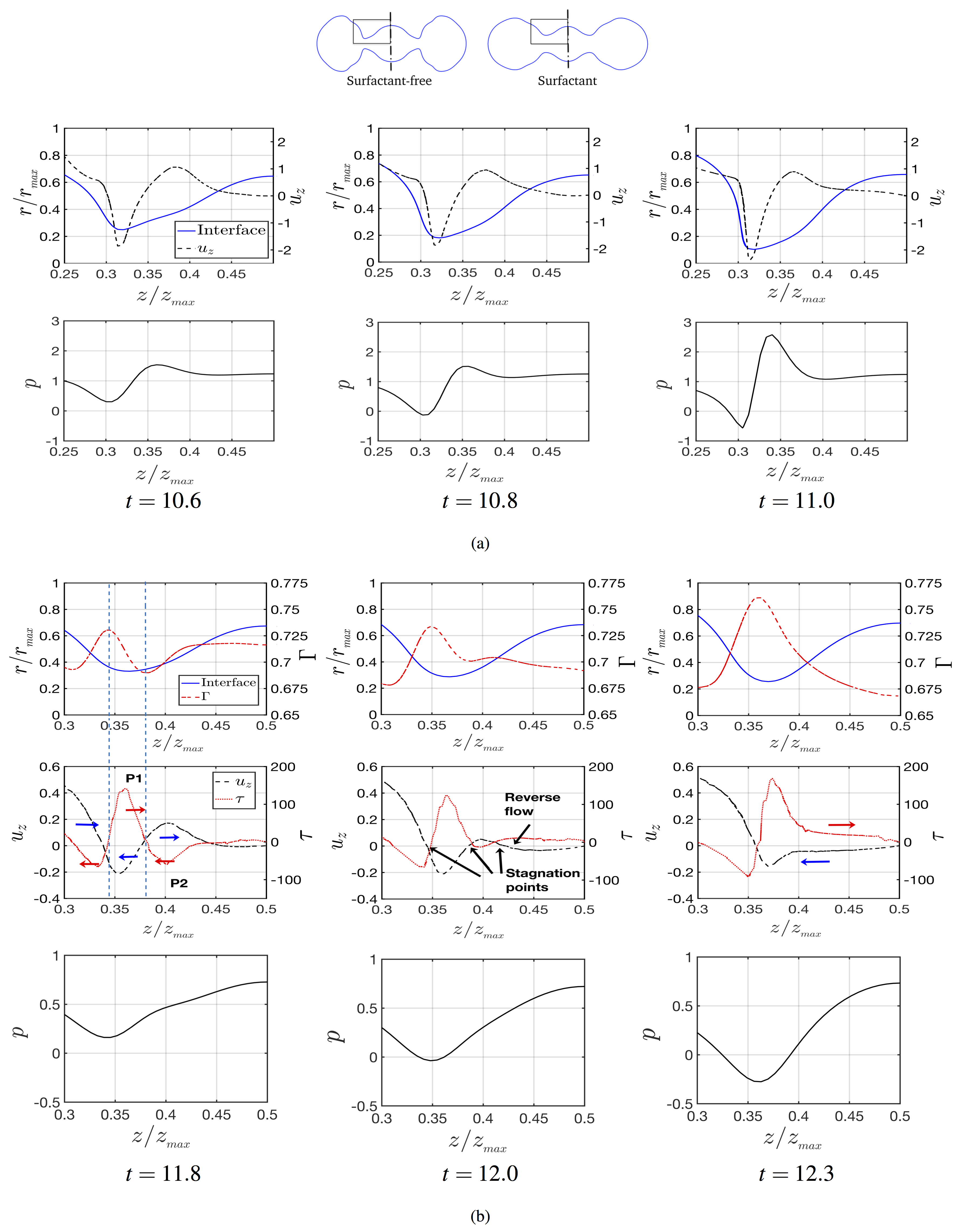}  
\caption{ \label{fig:escape_fields} Spatio-temporal evolution of the interfacial shape, $u_z$, and $p$ for the surfactant-free case, (a), and the surfactant-laden case, (b), for the same parameters as those used to generate Fig. \ref{fig:escape_vorticity}. In (b), we have also plotted the evolution of $\Gamma$ and $\tau$, 
and labelled the 
the direction of the  Marangoni-induced and capillary-driven flows with red and blue arrows, respectively. 
The dimensionless times are shown in each panel.}
\end{figure}

Figure \ref{fig:escape_fields}-(a) shows the coupling between the interface, $u_z$, and $p$ for the surfactant-free case as the neck evolves towards its capillary breakup. We have limited the fields to the framed region shown at the top of  Fig. \ref{fig:escape_fields}, due to the axisymmetric behaviour of the system. As the neck reduces its size, capillary pressure drives the flow, pumping fluid towards the neck, and, consequently, $u_z$ increases over time.

Similarly, Fig. \ref{fig:escape_fields}-(b) shows the coupling between $\Gamma$, $\tau$, $u_z$, $p$ and the interface for the surfactant-laden base case. At $t=11.8$, $u_z$ has a qualitatively similar behaviour to that associated with the surfactant-free case though the competition  between Marangoni stresses and the capillary pressure determines the direction of the flow. Inspection of $\tau$ at $t=11.8$ reveals the presence of two Marangoni peaks of opposite sign is observed; we have labelled the positive and negative peaks `P1' and `P2', respectively. We have also labelled with red and blue arrows the direction associated with Marangoni-induced and capillary-driven flow. 

In the region where P2 is located (see Fig. \ref{fig:escape_fields}-(b) at $t=11.8$), Marangoni stresses and capillary pressure drive flow towards the neck and the centre of the ligament, respectively. Hence, both mechanisms act in opposing directions. The induced $\tau$ decelerates the flow caused by the capillary pressure (observe that $u_z$ at $t=11.8$  and $t=12.0$ is equal to $0.18$ and $0.05$, respectively). At $t=12.0$, an additional stagnation point appears, as already discussed in connection with Fig. \ref{fig:escape_vorticity}-(b) and the deceleration continues until flow-reversal occurs at $t=12.1$ with the merging of two stagnation points. Furthermore, the $\tau$ peak P1 also decelerates the flow which passes through the neck though the neck size reduces over time in this case. After this point, the flow behaviour follows that described previously in the discussion of Fig. \ref{fig:escape_vorticity}-(b). It is also noteworthy that a comparison of the pressure $p$ for the clean and surfactant-laden cases reveals that a reduction of the capillary pressure is observed in the latter case due to the reduction in surface tension.

\begin{figure}
\includegraphics[width=1\linewidth]{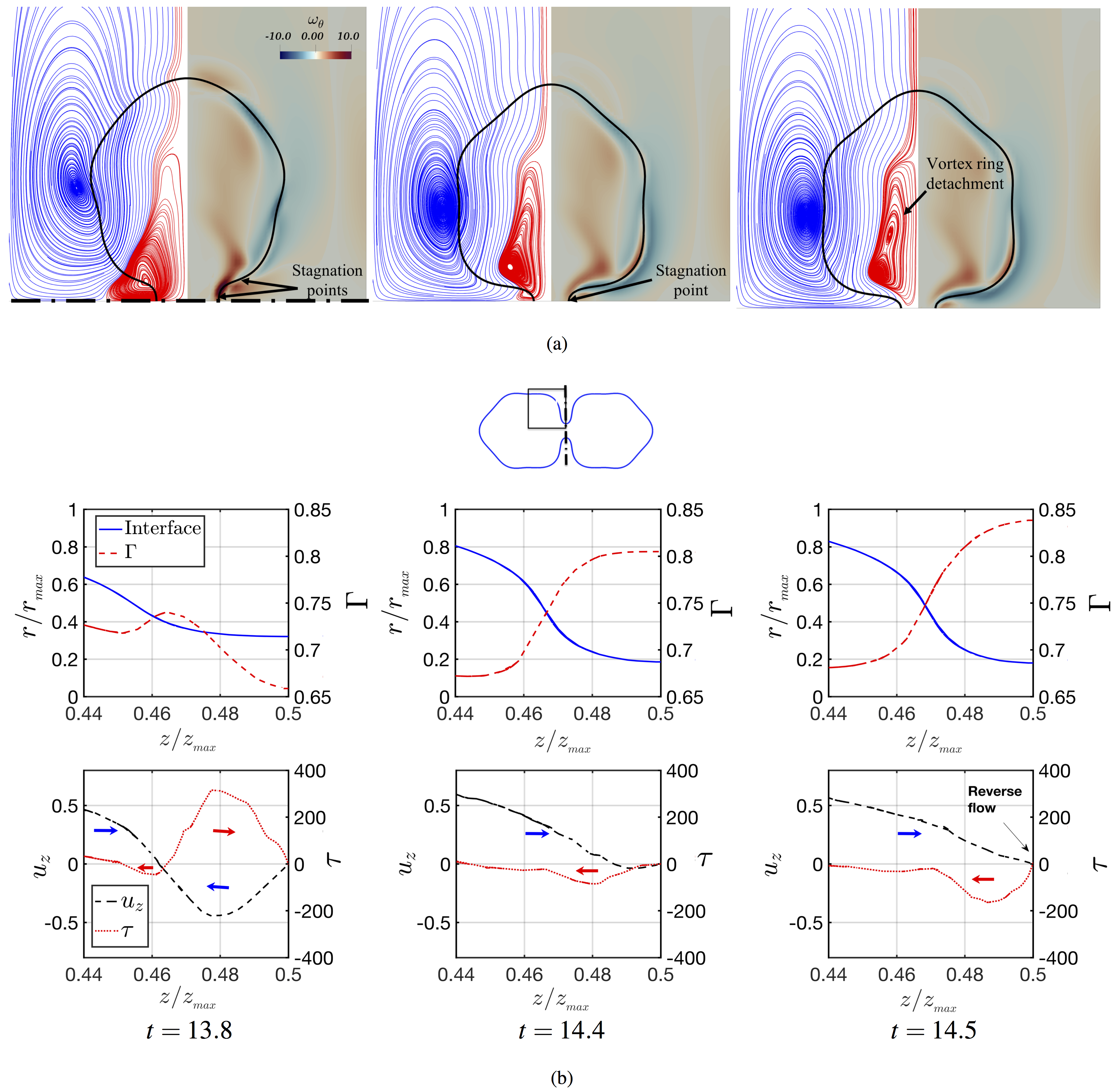}  
\caption{ \label{fig:escape_2}  Illustration of the surfactant-mediated mechanism underlying the second escape from pinchoff in Fig. \ref{fig:surfactant_base_case}-(c) depicting a three-dimensional representation of the velocity vector field, together with $\omega_{\theta}$ and streamlines, (a), at $t=13.8$, 14.4, and 14.5.
The framed regions of the ligament show a magnified view of the flow direction at each time. In (b), we have plotted the spatio-temporal evolution of the interfacial shape, $\Gamma$, $u_z$, and $\tau$ and labelled the direction of the  Marangoni-induced and capillary-driven flows with red and blue arrows, respectively. }
\end{figure}

Finally, we examine the flow field associated with the second reopening of the neck, $t \sim 14.5$, depicted in Fig. \ref{fig:surfactant_base_case}-(c). As shown at $t =13.8$ in Fig. \ref{fig:escape_2}, a large counter-clockwise rotating vortex is located at the neck, which drives flow from the neck towards the centre of the bulbous region as shown through the representation of the velocity vector fields. At $t =13.8$, the displacement of the vortical ring from the neck towards the bulk is observed. This vortex moves flow from the neck to the bulk triggering the reopening of the neck. Comparing the neck size at $t\ge14.8$, the neck does not stretch further and resists capillary breakup.

Figure \ref{fig:escape_2} shows simultaneously a snapshot of the interface, $\Gamma$, and $u_z$ at $t=13.8$ in which two stagnation points are observed; the highest value of $\Gamma$ is located close to the neck that links the neck with the bulbous region while its lowest value is at the centre of the ligament. Subsequently, Marangoni stresses  induce flow towards the second stagnation point (see the strong magnitude peak of $\tau$; a red arrow has been added to highlight the induced flow due to the concentration gradients).  In this region, the flow induced by $\tau$ competes with that driven by the capillary pressure in the opposite direction (shown with the blue arrow).
As time evolves, $\Gamma$ is convected towards the centre of the ligament which becomes 
a point of convergence and $\tau$ acts to oppose the surfactant accumulation (see $t=14.4$ and $t=14.5$); $\tau$ decelerates the flow induced by the capillary pressure and triggers flow-reversal, which is highlighted by the merging of the two stagnation points  at  $t=14.5$.   The flow reversal is  due to  the decrease in magnitude of the capillary pressure because of the surfactant-induced local reduction in surface tension; consequently, the capillary-driven flow is not strong enough to overcome that due to Marangoni stresses. With the suppression of one of the stagnation points, the vortex ring is displaced towards the bulk of the bulbous region.

%
%
%
\subsection{Parametric study}
Here, we investigate the fate of the ligament on system parameters such as the dimensionless elasticity parameter, $\beta_s$, the surface Peclet number, $Pe_s$, the Biot number, $Bi$, and the adsorption parameter, $\chi$. Unless stated otherwise, the parameters remain fixed to their `base' values: $L_o=15$, $Oh=10^{-2}$, $\beta_s=0.3$, and $Pe_s=10$. 
\begin{figure}
\includegraphics[width=1\linewidth]{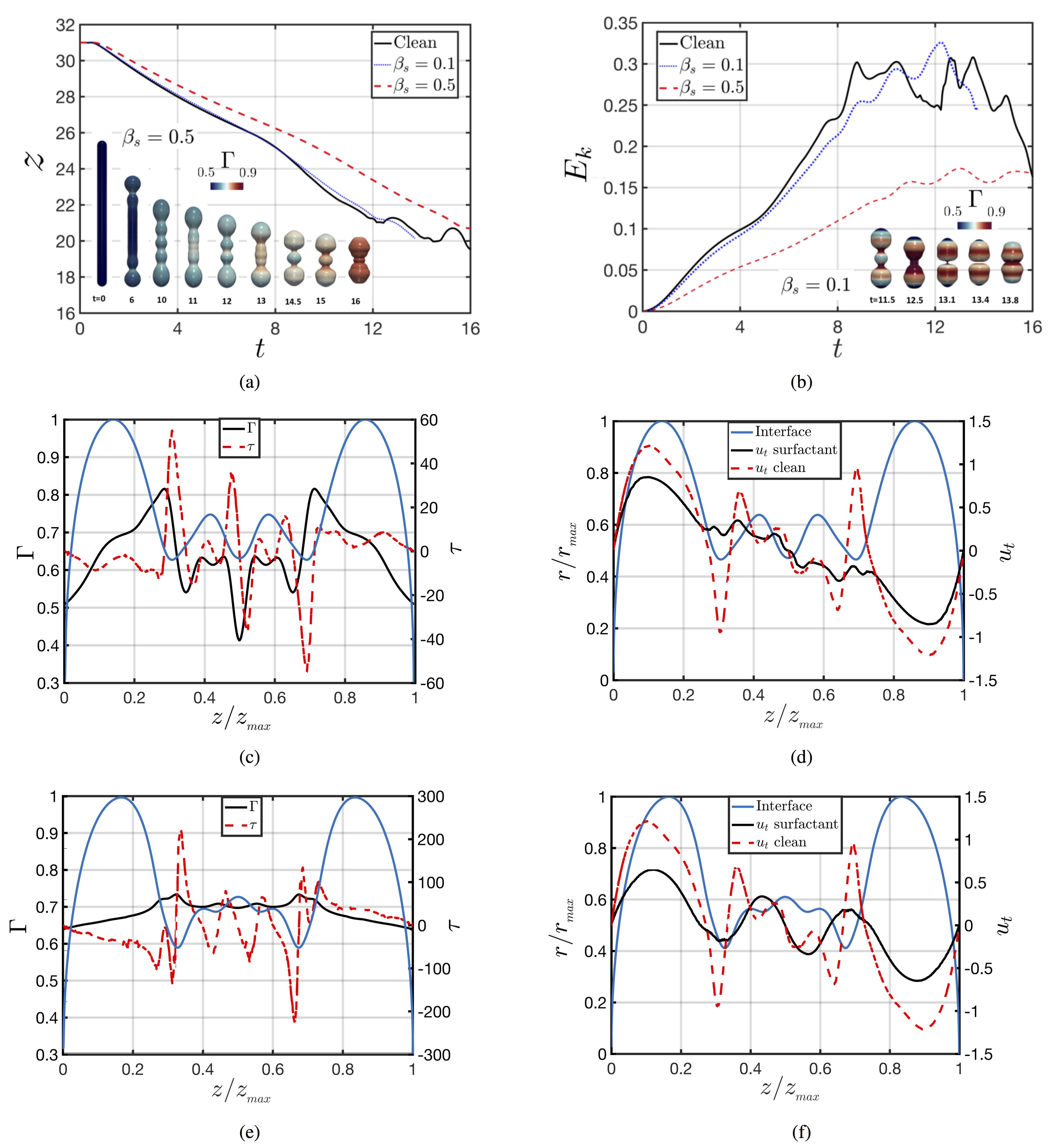}  
\caption{Effect of $\beta_s$ on the retraction dynamics for $L_o = 15$, $Oh=10^{-2}$, $Pe_s=10$, and $\Gamma_o =\chi\Gamma_\infty/2$: (a) and (b) temporal evolution of the north-tip location, and the kinetic energy, $E_k$, respectively; (a) and (b) also show three-dimensional representations of the interface for $\beta_s=0.5$ and 0.1, respectively, and for the dimensionless times shown in the panels in which the colour bar depicts the magnitude of the surfactant interfacial concentration, $\Gamma$. Panels (c) and (e), and (d) and (f) show the spatial variation of the interfacial shape, $\Gamma$, and $\tau$, and the tangential interfacial velocity, $u_t$, for the surfactant-free and surfactant-laden cases, respectively, at $t=10$. In (c) and (e), and (d) and (f), $\beta_s=0.1$ and 0.5, respectively.
        Note: in (c) interface location is superimposed, however, axis is not shown.\label{fig:marangoni_strenght}}
\end{figure}

\begin{figure}
\includegraphics[width=1\linewidth]{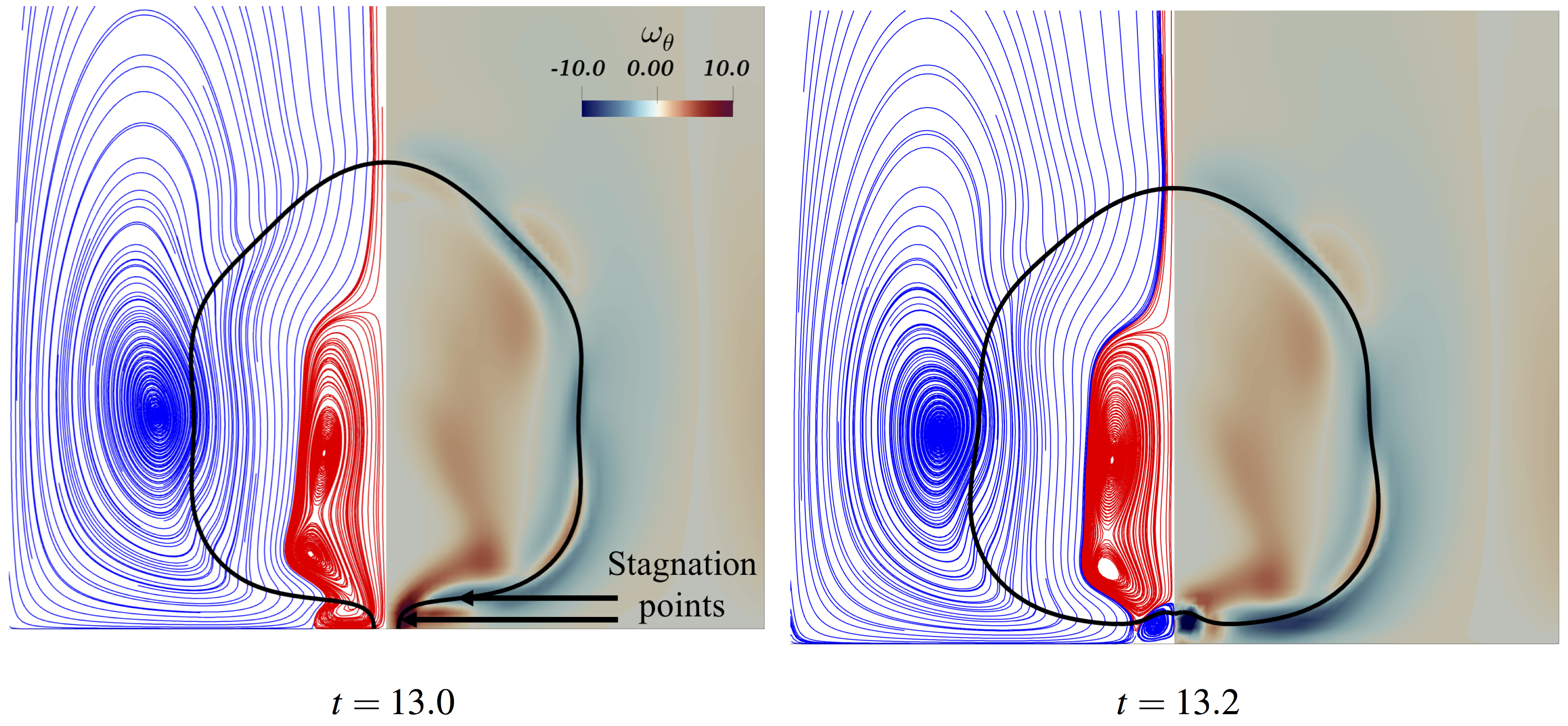}  
\caption{ \label{fig:no_escape}  Illustration of the no-escape from pinchoff for $\beta_s =0.1$ at $t=13$ and $13.2$. The rest of the parameters are the same as in Fig. \ref{fig:marangoni_strenght}.}
\end{figure}

We begin by examining the effect of parameter $\beta_s$, which characterises the relative significance of Marangoni stresses. 
As highlighted above, the redistribution of surfactant along the interface gives rise to concentration gradients and Marangoni stresses that act to retard retraction and prevent ligament pinchoff. Further evidence for this is provided in Fig. \ref{fig:marangoni_strenght}(a)-(b) in which we plot the temporal evolution of the ligament tip location, and the kinetic energy, respectively, for $\beta_s=0.1$ and $0.5$.
In Fig. \ref{fig:marangoni_strenght}, we also show a three-dimensional representation of the interface for these $\beta_s$ values. With increasing $\beta_s$, the Marangoni stresses are strengthened leading to a larger reduction in the retraction velocity and highlighting their retarding effect on the dynamics. 
As can also be seen clearly from  Fig. \ref{fig:marangoni_strenght}(a)-(b), for sufficiently large $\beta_s$ values, Marangoni stresses dominate the flow preventing ligament breakup.  
In panels (c) and (e), and (d) and (f) of Fig. \ref{fig:marangoni_strenght}, in which we plot a snapshot of the interfacial shape, $\Gamma$, $\tau$ and $u_t$ (for the clean and surfactant-laden cases), for $\beta_s=0.1$ and 0.5, respectively, it is shown that for the higher $\beta_s$ value, the larger Marangoni stresses lead to a  more uniform distribution of surfactant along the interface and a greater degree of interfacial rigidification; 
this is illustrated further through the overall reduction in $u_t$ and $E_K$ (see Fig. \ref{fig:marangoni_strenght}-(b)) with increasing $\beta_s$. 

For $\beta_s=0.5$, the escape from the end-pinching mechanism is observed as described in the previous section. For $\beta_s=0.1$, however, the relative importance of Marangoni stresses is smaller (see $\tau$ magnitude on Fig. \ref{fig:marangoni_strenght}-(c) which does not lead to prevention of the capillary breakup of the ligament (see interface at $t=13.4)$. Figure \ref{fig:no_escape} shows that $\tau$ is not sufficiently large to reverse the flow and the merging of the stagnation points does not occur here. Due to the existence of these two stagnation points close to the neck, the vortex ring close to the neck is not displaced towards the bulk (similar to Fig. \ref{fig:escape_2} at $t=14.4$), which is the genesis of the escape of pinchoff. Therefore, we are identifying  two different regimes with respect to $\beta_s$, and the transition region between the two regimes is located between $\beta_s=0.1$ and $\beta_s=0.3$.

\begin{figure}
\includegraphics[width=1\linewidth]{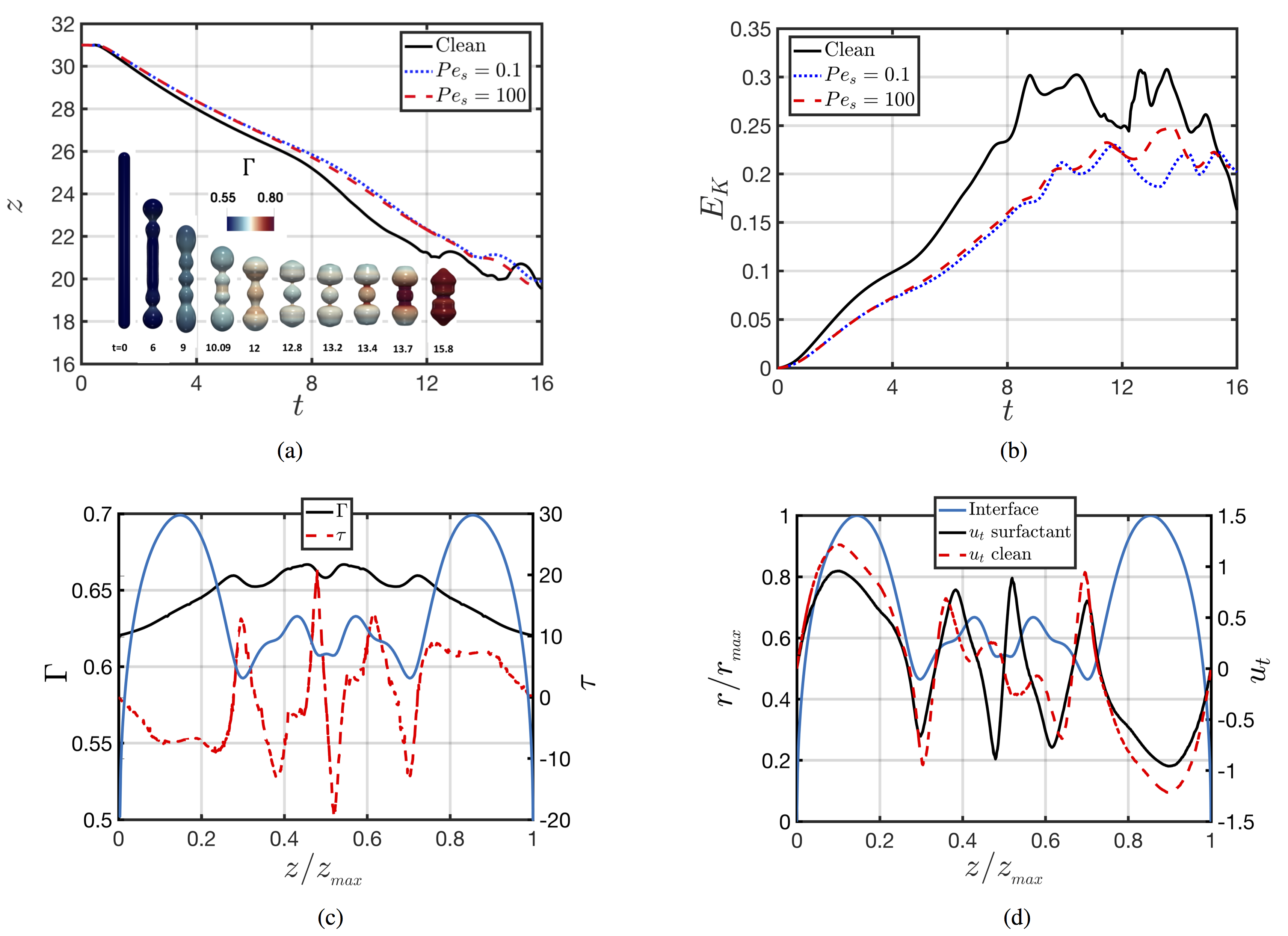}  
\caption{Effect of $Pe_s$ on the retraction dynamics for $L_o=15$, $Oh=10^{-2}$, $\beta_s=0.3$, $\chi=0.9$ and $\Gamma_o =\chi\Gamma_\infty/2$: (a) temporal evolution of the north-tip location and three-dimensional representations of the interface for $Pe_s=0.1$ and for the dimensionless times shown in the panels in which the colour bar depicts the magnitude of the surfactant interfacial concentration, $\Gamma$; (b) temporal evolution of the kinetic energy, $E_k$.  respectively; (a) and (b) also show  $\beta_s=0.5$ and 0.1, Panels (c) and (d) show the spatial variation of the interfacial shape, $\Gamma$, and $\tau$, and the tangential interfacial velocity, $u_t$, for the surfactant-free and surfactant-laden cases, respectively, for $Pe_s=0.1$ and at $t=10$.
        Note: in (c) and (e) interface location is superimposed, however, axis is not shown. \label{fig:pe_panel}}
\end{figure}

In Fig. \ref{fig:pe_panel}, we show the effect of varying $Pe_s$, which reflects the influence of surfactant diffusion effects along the interface, on the retraction dynamics. Inspection of this figure reveals that the promotion of diffusive effects through a decrease in $Pe_s$ leads to a more uniform interfacial distribution of $\Gamma$ and a reduction in the magnitude of surface tension gradients. It can also be seen that the retraction speed and ligament kinetic energy are weakly-dependent on $Pe_s$: they exhibit quantitatively similar dynamics over a three orders of magnitude variation in $Pe_s$. Therefore, keeping the other parameters fixed, the inhibition of the end-pinching mechanism (see \ref{fig:pe_panel}-(a) at $t=12.8$) is expected for cases when $Pe_s>0.1$.

\begin{figure}
\includegraphics[width=1\linewidth]{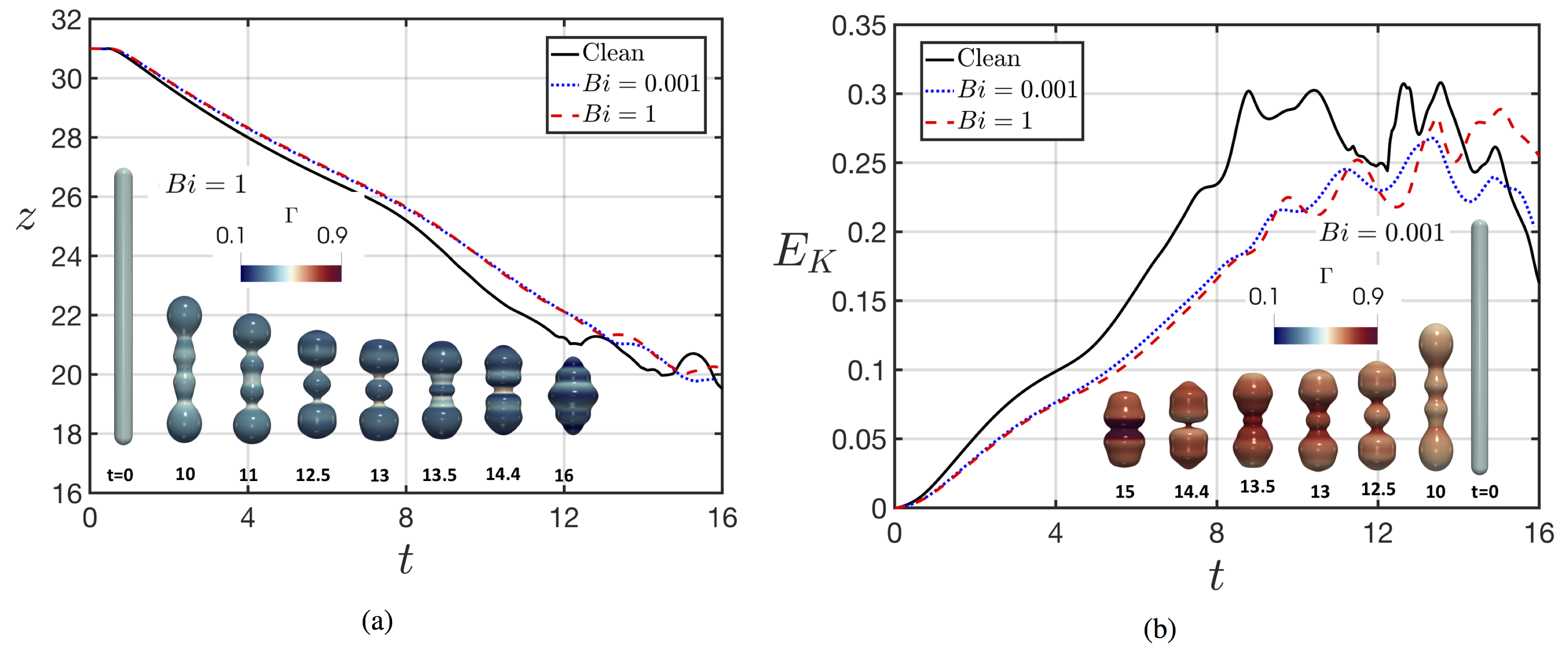}  
        \caption{Effect of $Bi$ on the retraction dynamics for $L_o = 15$, $Oh=10^{-2}$, $\beta_s=0.3$, $Pe_s=10$, $Pe_b=10$ and $\Gamma_o =\chi\Gamma_\infty/2$: (a) and (b) temporal evolution of the north-tip location, and the kinetic energy, $E_k$, respectively; (a) and (b) also show three-dimensional representations of the interface for $Bi=1$ and $10^{-3}$, respectively, and for the dimensionless times shown in the panels in which the colour bar depicts the magnitude of the surfactant interfacial concentration, $\Gamma$. Panels (c) and (e), and (d) and (f) show the spatial variation of the interfacial shape, $\Gamma$, and $\tau$, and the tangential interfacial velocity, $u_t$, for the surfactant-free and surfactant-laden cases, respectively, at $t=10$. In (c) and (e), and (d) and (f), $Bi=10^{-3}$ and 1, respectively.
        }
        \label{fig:soluble_bi}
\end{figure}

Up to this point, we have only analysed the fate of the ligament in presence of insoluble surfactants; here,  
we investigate the effect of surfactant solubility on the dynamics by fixing value of the fractional coverage to  $\chi=0.9$ and exploring the range $Bi=10^{-3}-1$. At the lower end of this range, the sorptive time scales are much larger than those associated with interfacial effects; consequently, the dynamics are dominated by capillarity and Marangoni stresses and are expected to be similar to those observed in the insoluble surfactant case.  For $Bi=O(1)$, the sorptive time scales are comparable to their capillary and Marangoni counterparts and the flow will reflect the delicate interplay amongst these effects. Inspection of Fig. \ref{fig:soluble_bi}-(a) and (b), however, shows that, contrary to expectations, $Bi$ has a relatively minor effect on the retraction speed and the ligament kinetic energy. From the three-dimensional representations of the interface, it can be seen that the ligament escapes its breakup for all $Bi$. For $Bi=1$, we observe the escape from breakup at $t \sim 12.4$ and $t \sim 14$. The radius of the neck prior to its escape also  increases with $Bi$.
Therefore, keeping the other parameters the same, the inhibition of the end-pinching mechanism (e.g. see \ref{fig:soluble_bi}-(a) at $t=12.5$)  is observed for three orders of magnitude of $Bi$.

\begin{figure}
\includegraphics[width=1\linewidth]{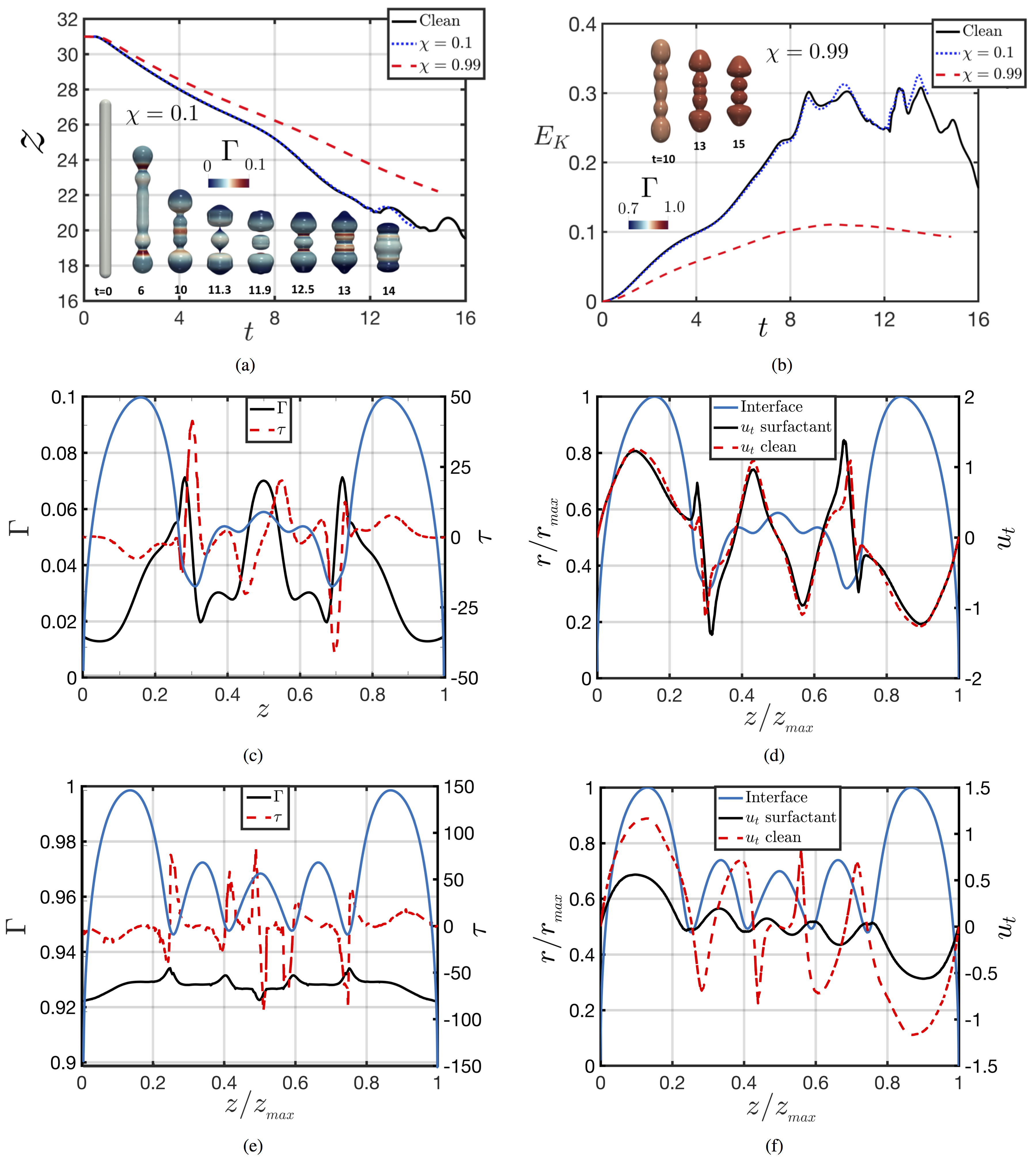}  
        \caption{Effect of $\chi$ on the retraction dynamics for $L_o = 15$, $Oh=10^{-2}$, $\beta_s=0.3$, $Pe_s=10$, $Pe_b=10$ and $Bi=0.1$: (a) and (b) temporal evolution of the north-tip location, and the kinetic energy, $E_k$, respectively; (a) and (b) also show three-dimensional representations of the interface for $\chi=0.1$ and $0.99$, respectively, and for the dimensionless times shown in the panels in which the colour bar depicts the magnitude of the surfactant interfacial concentration, $\Gamma$. Panels (c) and (e), and (d) and (f) show the spatial variation of the interfacial shape, $\Gamma$, and $\tau$, and the tangential interfacial velocity, $u_t$, for the surfactant-free and surfactant-laden cases, respectively, at $t=10$. In (c) and (e), and (d) and (f), $\chi=0.1$ and 0.99, respectively.
        Note: in (c) and (e) interface location is superimposed, however, axis is not shown.}
        \label{fig:panel_x}
\end{figure}
We now investigate the effect of the fractional coverage, represented by $\chi$ on the dynamics with $Bi=0.1$ and the rest of the parameters set to their base values.  Fig. \ref{fig:panel_x}-(a) and (b) shows that whereas the low $\chi$ dynamics resemble that of the surfactant-free case, at high $\chi$, for which adsorption effects are dominant, a significant reduction in the retraction velocity and kinetic energy is observed. 
Furthermore, as can be seen in Fig. \ref{fig:panel_x}-(c) and (f), for large $\chi$, interfacial gradients of the surfactant concentration, and therefore of surface tension, are small, which implies that Marangoni stresses play a minor role in this case. Thus, the reduction in ligament retraction velocity must be related to the significant reduction in surface tension, which acts to diminish the magnitude of capillary effects.
Therefore, for a low $\chi$, the relative importance of Marangoni stresses is very small (see $\tau$ magnitude in Fig. \ref{fig:panel_x}-c) which does not lead to the prevention of capillary breakup (see interface at $t=11.3)$.

\section{Conclusions\label{sec:conclusions}}

We have presented the effect of surfactant on ligament retraction of an aspect ratio $L_0 = 15$ and for intermediate Ohnersorge numbers, $Oh \sim 10 ^{-2}$. We have performed fully three-dimensional numerical simulations of the retracting process over a range of system parameters that account for the surfactant solubility and sorption kinetics and Marangoni stresses. The numerical method has been validated against the work of \citet{Notz_jfm_2004} for a surfactant-free case. Our results indicate that the presence of surfactant inhibits the end-pinching mechanism  and promotes the neck re-opening through Marangoni-flow, induced by the formation of surfactant concentration gradients, and not via lowering of the mean surface tension.
The induced Marangoni stresses decelerate the flow caused by the capillary pressure until flow-reversal occurs close to the neck. As the flow re-enters through the neck, the formation of a jet towards the bulbous region is observed, which gives rise to a vortex ring that eventually detaches towards the centre of the bulbous region. This behaviour is similar to the phenomenon explained by \citet{Hoepffner_jfm_2013} in the escape from breakup of viscous ligaments in the absence of surfactant. Therefore, the presence of surfactants avoids the `end-pinching' mechanism because of the existence of Marangoni stresses that suppress the mechanism of the Rayleigh-Plateau instability. This inhibition of the flow singularities is a remarkable outcome in the presence of a contaminant. We have also demonstrated that the pinchoff inhibition cannot happen by simply reducing the value of the surface tension, but only by the introduction of the Marangoni stresses. The presence of Marangoni stresses also leads to interfacial rigidification, which is observed through reduction of the ligament retraction velocity and the ligament kinetic energy. 
We have also investigated how the variation of key surfactant parameters affects the fate of the ligament. At $\beta_s=0.1$, Marangoni stresses are insufficiently large to influence the flow close to the neck and do not prevent the capillary breakup of the ligament. We also found that for the whole studied range of $Pe_s$ and $Bi$ an escape from end-pinching occurred. Additionally, we  showed that solubility, contrary to expectations, has a relatively minor effect on the retraction speed and the ligament kinetic energy. Finally, the adsorption effects were studied via variation of $\chi$, where at $\chi=0.1$, Marangoni stresses do not lead to the prevention of capillary breakup.

This research is of
importance for many applications that aim to produce equal-sized droplets, which is a desired outcome for improving efficiency in technologies such as ink-jet printing.
An interesting future line of research would be to study the one-dimensional free-surface slender cylindrical flow on the Navier-Stokes equations coupled with a set of equations describing the surfactant interfacial transport on a Newtonian liquid thread. This analysis could provide insights into the relationship between the classic Taylor-Culick retraction velocity and interfacial rigidification brought by the surface-active agents.
Future research avenues involve performing numerical simulations of curved ligaments to breakup the symmetry behaviour and of non-Newtonian ligaments, including visco-plastic and visco-elastic types, with a large range of Ohnesorge numbers and ligament aspect ratios. The retraction dynamics of non-axisymmetric ligaments may lead to the formation of `entrapped bubbles' inside the ligament and a three-dimensional oscillatory dynamics of the formed droplet.

\subsection*{Acknowledgements}

This work is supported by the Engineering \& Physical Sciences Research Council, United Kingdom, through a studentship for RC-A in the Centre for Doctoral Training on Theory and Simulation of Materials at Imperial College London funded by the EPSRC (EP/L015579/1) (Award reference:1808927), and through the EPSRC MEMPHIS (EP/K003976/1) and PREMIERE (EP/T000414/1) Programme Grants. 
OKM also acknowledges funding from PETRONAS and the Royal Academy of Engineering for a Research Chair in Multiphase Fluid Dynamics.
We also acknowledge the Thomas Young Centre under grant number TYC-101.  DJ and JC acknowledge support through computing time at the Institut du Developpement et des Ressources en Informatique Scientifique (IDRIS) of the Centre National de la Recherche Scientifique (CNRS), coordinated by GENCI
(Grand Equipement National de Calcul Intensif) Grant 2020 A0082B06721. The numerical simulations were
performed with code BLUE (\citet{Shin_jmst_2017}) and the visualisations have been generated
using ParaView.

\subsection*{APPENDIX}

\subsection*{Mesh study and numerical method}

Filament profiles (characterised by $L_o=15$ and $Oh=10^{-2}$) are compared against \citet{Notz_jfm_2004}, from which data was extracted by image analysis. They used a  Galerkin finite element approach under a  two-dimensional  axisymmetric  assumption in  the  azimuthal  direction.  As shown in Fig. 2, the ligament profiles  show a good qualitative agreement with our numerical method. The second check that we are performing is the predicted time for the capillary breakup for the surfactant-free case between the two methods (Notz and Basaran's pinchoff time is 11.114 and our pinchoff time for different meshes are shown in Table I), which shows a good quantitative agreement between the two methods. These two checks allow us to proceed with caution.

\begin{figure}
\includegraphics[width=0.8\linewidth]{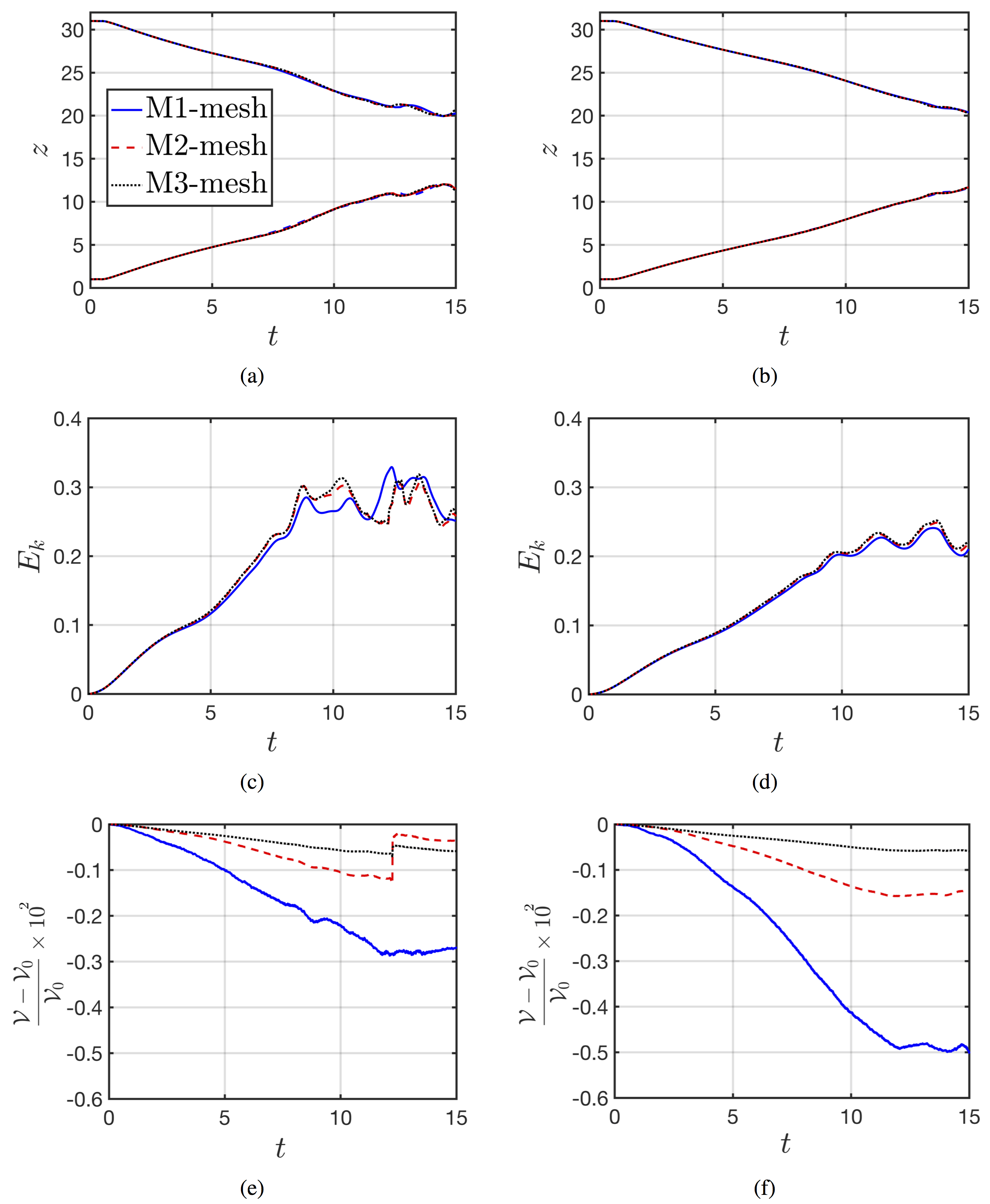}  
\caption{\label{fig:mesh} Mesh study for
`surfactant-free' (panels (a),(c), and (e) on the left) when $L_o = 15$ and $Oh=10^{-2}$, and `surfactant-laden base case' (panels (b), (d) and (f) on the right) when $L_o = 15$, $Oh=10^{-2}$, $\beta_s=0.3$ and $Pe_s=10$. The panels highlight the temporal evolution of the north and south tips location (a-b), the kinetic energy $E_k$ (c-d), and the relative variation of  the liquid volume (e-f).}
\end{figure}

The next question is to ensure that our numerical results are mesh-independent. To this end, the dynamics of the retracting ligament  for the surfactant free (i.e., $L_o=15$ and $Oh=10^{-2}$) and surfactant laden base cases (i.e.,  $L_o = 15$, $Oh=10^{-2}$, $\beta_s=0.3$ and $Pe_s=10$) are tested for different mesh resolutions (see Fig. \ref{fig:mesh}) in terms of the temporal evolution of the tips location, $E_k$ and the relative variation of  the liquid volume. The main characteristics of the meshes are summarised in Table \ref{tab:mesh_pinchoff} including the number of elements, and the predicted pinchoff time for the surfactant-free case.  As for the time $t \approx 12-12.5$, the resulting three droplets coalesce with each other to form a single droplet (shown in Fig. \ref{fig:mesh}-a). The volume of the liquid is conserved with a loss of less than $0.2\%$ during the topological changes. Because the curves for the kinetic energy and the location of the tips overlap for M2 and M3 meshes, we conclude that M2 is sufficiently refined to ensure mesh-independent results while providing a good compromise with the computational cost of the simulation. The results presented in this paper correspond to the M2 mesh. Additionally, We have carefully checked the accuracy of the surfactant mass conservation in all our simulations. Our numerical framework conserves surfactant moles with a relative error of $0.062\%$ for the surfactant-laden base case. Moreover, the code conserves surfactant moles with a relative error of $0.08\%$ (for case: $L_o = 15$, $Oh=10^{-2}$, $\beta_s=0.3$, $Pe_s=10$, $Pe_b=10$, $\chi=0.9$ and $Bi=0.1$).

\begin{table} [h!]
\centering
\caption{Retracting ligament mesh study for the surfactant free case ($L_o=15$ and $Oh=10^{-2}$).}
\begin{tabular}{cccc}
~ Run ~  & \begin{tabular}[c]{@{}l@{}}~ Global mesh size\\~ (number of cells) ~ ~\end{tabular} & \begin{tabular}[c]{@{}l@{}}~ Number of parallel ~\\~ ~process threads ~ ~\end{tabular} & \begin{tabular}[c]{@{}l@{}}~ Pinchoff time\\~~~~~~~~~~(s) ~ ~\end{tabular}   \\
~ M1 ~  & $96\times96\times384$        & $54$   & $11.1151$         \\
~ M2 ~     & $192\times192\times768$        & $432$ &  $11.1158$   \\
~ M3 ~     &$384\times384\times1536$          & $3456$   & $11.1167$        
\end{tabular}
\label{tab:mesh_pinchoff}
\end{table}

The temporal integration scheme is based on a second-order Gear method (\citet{Trucker_Springer_2013}), with implicit solution of the viscous terms of the velocity components. Each time step is computed by an adaptive time step-size criterion:

\begin{equation}
\Delta t = \min \left\{\Delta t_{cap},~\Delta t_{vis},~\Delta t_{_{CFL}}, ~\Delta t_{int}\right\}
\end{equation}

\noindent

where $\Delta t_{cap},~\Delta t_{vis},~\Delta t_{_{CFL}},$ and$~\Delta t_{int}$ represent the capillary time step, the viscous time step, the Courant- Friedrichs-Lewy (CFL) time step, and interfacial CFL time step, respectively. Those terms are defined by

\begin{equation} 
\left.\begin{array}{ccc}

\displaystyle{\Delta t_{cap} =\frac{1}{2} \sqrt{\frac{(\rho_{_l} +\rho_{_g}) \Delta {x}^3}{\pi \sigma_s }}},& ~& \displaystyle{\Delta t_{vis} =\frac{\rho_{_g} \Delta x^2 }{6 \mu_{_l} }} \\
\displaystyle{\Delta t_{_{CFL}} =\frac{ \Delta x }{|u_{max}|}},& ~&  \displaystyle{\Delta t_{int} =  \frac{ \Delta x }{\left | V_{int} \right |} }
\end{array}\right.
\end{equation}

\noindent
where $\Delta x$ refers to the minimum cell size, $u_{max}$ and $V_{int}$ are the maximum fluid and interface velocities, respectively. More details of the numerical method are provided in \citet{Shin_jcp_2018,Shin_jmst_2017}. A brief summary of the numerical aspects that are relevant to the study is presented here. The Navier Stokes equations are solved by a finite volume method on a staggered grid (\citet{Harlow_pof_1965}). The computational domain is discretised by a fixed regular grid (i.e. Eulerian grid) and the spatial derivatives are approximated by standard centred difference discretisation, except for the non-linear term, which makes use of a second-order essentially non-oscillatory (ENO) scheme \cite{Shu_jcp_1989,Sussman_cp_1998}. In the case of the viscous term, a second-order centred difference scheme is used. The projection method is used to treat the incompressibility condition \cite{Chorin_mc_1968}. A multigrid iterative method is used for solving the elliptic pressure Poisson equation  

With respect to the treatment of the free surface. The interface is tracked with an additional Lagrangian grid by using the Front-Tracking method together with the reconstruction of the interface by using the Level Contour Reconstruction Method  \citet{Shin_jcp_2002,Shin_ijnmf_2009}. Communication between the Eulerian and Lagrangian grid (for the transfer of the geometric information of the interface) is done by using the discrete delta function and the immersed boundary method of \citet{Peskin_jcp_1977}. The advection of the Lagrangian interface is done by  integrating $d\textbf{x}_f /dt = \textbf{V}$ with a second-order Runge-Kutta method, where $\textbf{V}$ stands for the interfacial velocity which has been calculated by interpolation from the Eulerian velocity. The code is parallelised through an algebraic domain-decomposition technique and the communication between subdomains for data exchange is managed by the Message Passing Interface (MPI) protocol. Figure \ref{fig:intro}-(b) highlights the partitioning of the computational domain into $6 \times 6 \times 12 = 432$ sub-domains.

Finally, due to the nature of the system, the surfactant transport is solved on the interface, where the surfactant concentration is located on the centre of the triangular front elements. In the convective-diffusive equations, the time is integrated by a first order explicit scheme. The surface surfactant gradients in the interface are computed by using a probing technique introduced by \citet{Udaykumar_jcp_1999}. A  Sharp boundary condition for bulk surfactant concentration equation is implemented to treat the source term at the interface.


 \end{document}